\documentclass{article}

\usepackage{arxiv}

\usepackage[utf8]{inputenc} 
\usepackage[T1]{fontenc}    
\usepackage{hyperref}       
\usepackage{url}            
\usepackage{booktabs}       
\usepackage{amsfonts}       
\usepackage{nicefrac}       
\usepackage{microtype}      
\usepackage{lipsum}

\usepackage{graphicx}
\usepackage{algorithm}
\usepackage{algpseudocode}
\usepackage{amsmath,amssymb,amsfonts}
\usepackage{textcomp}

\usepackage[table]{xcolor}
\definecolor{light-gray}{gray}{0.95}
\definecolor{dark-gray}{gray}{0.3}

\usepackage{amsthm}

\newcommand\concat{\mathbin{+\mkern-10mu+}}

\title{Group-Testing on Hypergraphs with \\ Variable-Cost Tests: A Power Systems Case Study}

\author{
  Laurence~A.~Clarfeld\thanks{Corresponding author: Laurence A. Clarfeld (e-mail: Laurence.Clarfeld@uvm.edu).} \\
  Department of Computer Science\\
  University of Vermont\\
  Burlington, VT, 05405 \\
  \texttt{Laurence.Clarfeld@uvm.edu} \\
   \And
 Margaret~J.~Eppstein \\
  Department of Electrical Engineering\\
  University of Vermont\\
  Burlington, VT, 05405 \\
  \texttt{Maggie.Eppstein@uvm.edu} \\
}

\begin{document}
\maketitle

\begin{abstract}
Assessing risk of cascading failure in an electrical grid requires identifying many small ``defective'' subsets of the $N$ elements in a power system, such that the simultaneous failure of all elements in a defective set triggers a large cascading blackout. Most supersets of defective sets will also initiate  cascades. This property was leveraged by Random Chemistry (RC), a state-of-the-art algorithm for efficiently finding minimal defective sets. While not previously acknowledged as such, RC is a group-testing algorithm, in that it tests pools of candidate solutions. RC then reduces the size of defective pools until a minimal defective set is found. Using a synthetic model of the Western United States power grid, we show that run times are minimized with a much smaller initial pool size than previously recommended, nearly doubling the speed of RC.  We compare RC to a proposed new alternative group-testing algorithm (SIGHT). Both algorithms have their advantages; while SIGHT is shown to require fewer total tests than RC, RC requires fewer tests of defective sets. Thus, the relative speed of each algorithm depends on the relative cost of defective \textit{vs.} non-defective tests, a factor not previously considered in the group-testing literature. In the power systems application, tests of defective sets are much more computationally costly, so RC is faster. However, with only a single control parameter (compared to many in RC), SIGHT is easier to optimize, and in applications where the cost ratio of defective:non-defective tests is low enough, SIGHT will be faster than RC. 
\end{abstract}


\section{Introduction}
\label{sec:introduction}
On August 14, 2003, a computer glitch allowed the failure of several transmission lines in Ohio, due to poor vegetation management, to trigger a cascading blackout that resulted in 45 million people losing electricity across the Northeastern United States and Canada \cite{muir2004final}. While large cascading power failures such as this are rare, due to their vast size, the risk associated with such events is significant \cite{chen2006probability, dobson2007complex, newman2011exploring}. Power systems are thus required to operate such that the failure of any single component will not trigger a large cascade. Grid operators are now also required to consider the risk associated with sets of $k$ elements of the power grid ($k>1$) failing together \cite{nerc2007top}. However, this is an extremely challenging task and there is no guarantee built into the system that two or more elements failing simultaneously will not cause a large blackout. 

Given a power grid with $N$ elements, analysis of the risk of cascading failure may be performed by running a simulation (a.k.a. test) to see if the failure of $k$ elements (a.k.a. $k$-set) results in a large blackout. If a large blackout results, the given $k$-set is considered ``defective''. In the power systems literature, this is often referred to as a blackout-causing $N-k$ contingency, but to be consistent with the more general group-testing literature we refer to these as defective $k$-sets in the remainder of this paper. For even modestly sized power grids, the search space of $k$-sets for $k>1$ is sufficiently large that it would be inefficient (for $k=2$) or infeasible (for $k>2$) to test them all. Thus, finding efficient methods for sampling defective $k$-sets is an important challenge. Because the superset of any defective set is also likely to trigger a cascade \cite{rezaei2014estimating}, the task of identifying minimal defective $k$-sets can be tackled as a group testing problem.

The field of group testing is thought to have originated from a single report by Dorfman in which he proposed a novel method for efficiently screening soldiers for syphilis during World War II \cite{dorfman1943detection}. Dorfman suggested mixing together blood samples from multiple individuals so it would require just a single chemical test to determine if the pooled blood sample contained syphilitic antigen. If the test came back negative, it would indicate that none of the soldiers were infected, whereas a positive test result would require subsequent tests to determine which soldiers were infected. Finding optimal group testing strategies that minimize the number of tests required to identify ``defective'' individuals (or items, or sets of items) has been a central focus in group testing research.  In this paper, we consider so-called ``adaptive'' group testing algorithms, where the tests in each stage of the algorithm depend on results of the previous stages. 

Searching for defective edges in a graph $G$ with $|E|$ edges using group testing was first considered by Aigner, who conjectured that no more than $\lceil\log_2|E|\rceil+c$ tests (for some constant $c$) were required to find a single defective edge in $G$ \cite{aigner1988combinatorial}. This was later proven by Damaschke \cite{damaschke1994tight} and generalized to hypergraphs by Triesch \cite{triesch1996group}. The case of finding $d>1$ defective edges in a graph, where $d$ is known, was first addressed in \cite{johann2002group}. For the case when $d$ is unknown, adaptive methods for finding all defective edges were proposed for graphs in \cite{hwang2005competitive} and extended to hypergraphs in \cite{chen2007competitive,chen2011revised}. 

Cascading failure in the power grid can be framed as a group testing problem using hypergraphs. A hypergraph is a generalization of an ordinary graph, in which individual edges (so-called ``hyperedges'') can connect an arbitrary number of nodes. Specifically, define a hypergraph $G(V,E)$ where each node in $V$ represents one of the $N$ elements of interest in a power system and each hyperedge in $E$ connects the nodes of a $k$-set, and there exists an edge in $E$ for every possible subset of $V$ with $k>1$.  The aim is to identify minimal defective $k$-sets in $E$ (for $k_{min} \leq k \leq k_{max}$), where ``minimal'' means that no smaller subset is defective (\textit{i.e.}, causes a cascading failure). 

When considering the application of group testing to identify defective $k$-sets that trigger cascading failures in power systems, there are several important distinctions from most previous work that has applied group testing to hypergraphs: (i) The number of minimal defective hyperedges is so large that it is not feasible to identify them all; (ii) $G$ is non-uniform (\textit{i.e.}, $k$ is variable, in the $k$-sets defined by the hyperedges in $E$); (iii) tests can produce false negatives, wherein a set tests as non-defective even though it contains a defective subset, (\textit{e.g.}, this can occur when the grid is fragmented into disconnected but independently functioning ``islands'' \cite{aghamohammadi2012intentional}); and, importantly, (iv) not all tests have the same cost (required computation time). 

The implication of (i) is that, rather than attempt to find all defective sets (as in \cite{johann2002group,hwang2005competitive,chen2007competitive,chen2011revised}), we seek a computationally efficient method for collecting a large sample of minimal defective $k$-sets. These samples can be used to identify which elements of a power system, when failed, are most likely to interact with other failures to trigger a large cascade, and can thus inform effective cascade prevention and/or mitigation strategies \cite{rezaei2014estimating,rezaei2015rapid}. In addition, these samples can be used to obtain estimates of the total number of minimal defective sets of each specified size $k$ (\textit{e.g.}, using methods such as in \cite{clarfeld2018assessing}), which are necessary for approximating system-wide risk of cascading failure \cite{rezaei2014estimating,rezaei2015rapid,clarfeld2018assessing}. The consequence of (ii) is that minimal defective $k$-sets can be arbitrarily large, and since (for practical reasons \cite{clarfeld2018assessing}) one may only be interested in identifying $k$-sets up to a certain maximum size $k_{max}$, the sampling method must allow for $k_{max}$ to be specified, and yet still be robust to the existence of defective $k$-sets where $k>k_{max}$. Due to (iii), the algorithm must be as robust as possible to the presence of false negatives. Finally, when the computational cost of tests is variable (iv), the number of required tests is not necessarily a direct proxy for the required running time of each algorithm. This is the case in the power systems application, where defective tests (simulations that find a defective $k$-set) have significantly longer running times than non-defective tests, as shown in Sec. \ref{sec:results}. 

Monte Carlo (MC) sampling can be used for finding defective $k$-sets in power systems \cite{allan2013reliability}, by simply testing randomly selected $k$-sets, where the elements of these sets are stochastically selected according to their joint probability of failure. MC sampling does not require $k_{max}$ to be pre-specified, but also does not guarantee that found defective $k$-sets are minimal. In addition, MC is computationally inefficient. Other methods accelerate the discovery of defective sets by leveraging the differential likelihoods of various elements being involved in minimal defective $k$-sets (\textit{e.g.}, \cite{donde2008severe,ma2018fast}). One disadvantage to these ``importance sampling'' approaches is that they bias against including elements that have low \textit{a priori} likelihoods of being involved in minimal defective sets. Since cascading failure is a highly non-linear process, this under-sampling can severely limit the space of all possible cascading failures that one samples from, and thus limit the insight that one gets from the data that result. 

In contrast, the Random Chemistry (RC) algorithm \cite{eppstein2012random} uses a ``top-down'' approach (moving from larger to smaller defective $k$-sets) to efficiently find minimal defective $k$-sets for $k \le  k_{max}$ in power grids. While not previously recognized as such, RC is effectively a stochastic adaptive group testing approach.  RC sampling has been found to be orders of magnitude faster than MC sampling, even on relatively small ($N < 3000$) power system models \cite{rezaei2014estimating}, and has been incorporated into state-of-the-art  algorithms designed to efficiently estimate the risk of cascading failure in power systems \cite{rezaei2014estimating, rezaei2015rapid, clarfeld2018assessing, clarfeld2019risk}. But while \cite{eppstein2012random} offered some heuristic guidelines for setting the many parameters required by RC, they did not claim to have optimized these heuristics.

Here, we compare RC sampling to a proposed new algorithm we refer to as SIGHT (Sampling Inspired by Group Hyperedge Testing) on the power systems application, using a large test case modeled after the Western United States power grid. SIGHT is a modification of traditional group testing strategies \cite{triesch1996group,hwang2005competitive,chen2007competitive}, where the aim is to minimize the number of required tests. Each run of SIGHT first stochastically chooses elements of an initial $k$-set to test and, if a defective initial $k$-set is found, deterministically seeks a single minimal defective $k$-set from within the initial set. The SIGHT algorithm offers some potential advantages over RC sampling in that (i) it has only one control parameter to optimize, and (ii) it requires fewer tests than RC to find a minimal defective $k$-set. To ensure a fair comparison, a parameter sweep was first conducted to determine the initial set size that minimized median run time on the test case for each algorithm.

Our experiments yielded some surprising and useful results. Specifically, we find that (i) the prior heuristics used for setting the initial set size in RC \cite{eppstein2012random, clarfeld2018assessing} were far from optimal, and (ii) although SIGHT does outperform RC in terms of the number of required tests, RC outperforms SIGHT in terms of run time on the power systems test case.  We discuss the circumstances under which SIGHT is expected to be faster than RC, and \textit{vice versa}. 

This paper is organized as follows: The proposed SIGHT algorithm is presented in Section \ref{sec:SIGHT} and the RC algorithm is reviewed in Section \ref{sec:RC}; Section \ref{sec:alg_opt} discusses parameterization of both group sampling algorithms; Experiments to compare the performance of the two algorithms for finding minimal defective $k$-sets that trigger cascading failure in a relatively large power systems test case are described in Section \ref{sec:experiments};  Results, Discussion and Conclusions are presented in Sections \ref{sec:results}, \ref{sec:discussion}, and \ref{sec:conclusion}, respectively.

\section{Methods}

\subsection{SIGHT} \label{sec:SIGHT}
\subsubsection{Group Testing Inspiration}
In the simple case of a single minimal defective edge in a graph, Triesch proposed a group testing halving procedure \cite{triesch1996group} that was later adapted by Johann \cite{johann2002group} to find all $d$ defective edges (2-sets) in a simple graph in at most $d(\lceil \log_2\frac{m}{d}\rceil+7)$ tests, where $d$ is known. The algorithm presented by Johann for finding a single minimal defective edge in a subgraph $G' \subset G$ has subsequently been referred to as the TJ Procedure \cite{hwang2005competitive,chen2007competitive} and is the primary inspiration for the $BinSearchSIGHT$ subroutine presented in Sec. \ref{sec:SIGHT_alg}. The TJ procedure was first used to find all $d$ defective edges, when $d$ is unknown, in simple graphs \cite{hwang2005competitive}, and then extended to hypergraphs \cite{chen2007competitive, chen2011revised}.  SIGHT is a new variation of the TJ procedure, modified in three ways: (i) it is designed to find a defective $k$-set for $k_{min} \leq k \leq k_{max}$, which differs from past implementations that assume all defective sets of all sizes will be found; (ii) it is designed to be robust to the presence of minimal defective sets where $k>k_{max}$; and (iii) it is designed to be robust to false negative tests. 

\subsubsection{SIGHT Algorithm} \label{sec:SIGHT_alg} 
SIGHT takes as input the universal set $V$ of nodes (\textit{e.g.}, all elements in a system to be considered as candidates for failure),  an initial subset size $a_0$, and the bounds $k_{min}$ and $k_{max}$ on the size of the defective $k$-sets one is searching for. It returns either a minimal defective $k$-set (for $k_{min} \leq k \leq k_{max}$) or the empty set (if the algorithm aborts). 

Since power networks are already operated such that there are no defective 1-sets, $k_{min}=2$ for this application.  For large power grids, $k_{max}$ is typically limited to 3 (or possibly 4), due to the sheer number of these $k$-sets \cite{clarfeld2018assessing}. The only control parameter required by SIGHT is $a_0$, which establishes the size of the initial subset to be tested, where $k_{max} \leq a_0 \leq N$. The algorithm is illustrated by the Psuedocode in Algorithm \ref{alg:SIGHT} and is described below. 

The first step of the SIGHT algorithm  (Alg. \ref{alg:SIGHT}: line 1) calls the subroutine $SAMPLE(V,a_0)$ (pseudocode not shown), which initializes a list $S$ to a uniform random sample, in random order, of $a_0$ unique elements from $V$. In the following description, the notation $S[i]$ is used to refer to the $i^{\mathrm{th}}$ element of $S$; $S[i]$ is considered to be left of $S[j]$ for all $i<j$; and $S[{\;:\,}i]$ refers to the first $i$ elements of $S$.

A list $D$ is initialized to the empty list (Alg. \ref{alg:SIGHT}: line 2). This list $D$ will be used to accumulate known nodes from defective $k$-sets.

\newpage

\begin{algorithm}
\caption{SIGHT$(V,a_0,k_{min},k_{max})$} \label{alg:SIGHT}
\begin{algorithmic}[1]
    \State $S\gets SAMPLE(V,a_0)$
    \State $D\gets \emptyset$
    \If {$\neg \, isDefective(S)$}
        \State \textbf{return} $\emptyset$
    \EndIf
    
    \While {$|D| < k_{max}$}
        \State $m \gets BinSearchSIGHT(S,D)$
        \State $D\gets D \concat S[m]$
        \If {$|D| \geq k_{min} \; \land isDefective(D)$ } \footnotemark
            \State $D \gets bottomUpSIGHT(D,k_{min},k_{max})$
            \State \textbf{return} $D$
        \EndIf
        
        \State $S\gets S[{\;:\,}m-1]$
    \EndWhile
    \State \textbf{return} $\emptyset$
\end{algorithmic}
\end{algorithm}
\footnotetext{The use of short-circuiting logic in the conditional statement prevents unnecessary tests.}

In Alg. \ref{alg:SIGHT}: line 3, the subroutine $isDefective(S)$ is called to test whether the list $S$ is a defective set (although not necessarily minimal). This algorithm is specific to the application domain, so is not shown here. In the power systems application, it comprises a power system simulator (such as \cite{Hines2016cascadingChapter}) that detects whether or not a cascading failure would occur if all elements in $S$ fail; typically, there is some threshold that specifies some minimum size or impact of a cascade that is considered  to be a  ``cascading failure''. The subroutine $isDefective(S)$ is \textit{expected} to return TRUE if there is a minimal defective $k$-set in $S$ and FALSE otherwise. However, in the power systems application, there is a small, unknown, probability of false negatives; \textit{i.e.}, where $isDefective(S)$ returns FALSE even when $S$ contains a minimal defective set, because in some cases power flow is rerouted after removing additional element(s) from a system in such a way that a large cascade is avoided \cite{rezaei2015rapid} (as taken advantage of in ``intentional islanding''~\cite{aghamohammadi2012intentional}). 

If $isDefective(S)$ returns FALSE, the algorithm aborts and returns the empty set (Alg. \ref{alg:SIGHT}: line 4). Otherwise, the algorithm calls the subroutine $BinSearchSIGHT(S,D)$ (Alg. \ref{alg:SIGHT}: line 7)

The subroutine $BinSearchSIGHT(S,D)$ (Algorithm \ref{alg:bin}), following \cite{johann2002group,chen2007competitive,hwang2005competitive},  uses a binary search to find the index of the leftmost element in $S$ that is the rightmost element of a defective $k$-set in $D \concat S$, where $\concat$ denotes list concatenation. The subroutine is implemented like a classic binary search, in that it first assigns left and right pointers $l$ and $r$ to the leftmost and rightmost elements of $S$, respectively (Alg. \ref{alg:bin}: lines 1-2), and determines a test index, $i$, half way between $l$ and $r$. But unlike a standard binary search that assumes that the values of the elements of $S$ are in sorted order and performs a test on element $S[i]$, here only the ordering of the indices matters (the actual elements in $S$ are deliberately in random order to prevent the algorithm from biasing towards sets that include elements with low values)  and the test (Alg. \ref{alg:bin}: line 5) includes all elements of $S$ with indices $\le i$ as well as all elements from the growing list $D$, to see whether $D \concat S[{\;:\,}r-i]$, is defective. If so, $r$ is reduced (Alg. \ref{alg:bin}: line 6); if not, $l$ is increased (Alg. \ref{alg:bin}: line 8). The process repeats until $l$ and $r$ converge to some index, which is returned. 

\begin{algorithm}
\caption{BinSearchSIGHT$(S,D)$}  \label{alg:bin}
\begin{algorithmic}[1]
    \State $l \gets 1$
    \State $r \gets |S|$
    \While {$l<r$}
        \State $i \gets \lceil \frac{r-l}{2}\rceil$
        \If {$isDefective(D \concat S[{\;:\,}r-i])$}
            \State $r \gets r-i$
        \Else
            \State $l \gets r - i + 1$
        \EndIf
    \EndWhile
    \State \textbf{return} r
\end{algorithmic}
\end{algorithm}

~\\

After each call to the $BinSearchSIGHT$ subroutine, SIGHT concatenates the found element $S[m]$ to $D$ (Alg. \ref{alg:SIGHT}: line 8). If the accumulated set $D$ is found to be defective and is at least size $k_{min}$ (Alg. \ref{alg:SIGHT}: line 9), then the algorithm has successfully found a list $D$ that contains a minimal defective $k$-set. However, because in the power systems application not all supersets of a defective set will cause a cascade, in some cases $D$ will be non-minimal. Thus, it is necessary to call the subroutine $bottomUpSIGHT(D,k_{min},k_{max})$ (Alg. \ref{alg:SIGHT}: line 10, pseudocode for $bottomUpSIGHT$ not shown), which tests any subsets of $D$ of size $k_{min} \le k \leq k_{max}$ that have not already been tested, to ensure that the defective subset returned is minimal  (Alg. \ref{alg:SIGHT}: line 11). If $D$ is not defective, then $S[m]$ and all elements to its right are removed from $S$ (Alg. \ref{alg:SIGHT}: line 13) and the process is repeated until either a defective is found, or $|D|$ exceeds $k_{max}$ (Alg. \ref{alg:SIGHT}: line 6), in which case the defective set being found is too large and the algorithm aborts (Alg. \ref{alg:SIGHT}: line 15). Note that Alg. \ref{alg:SIGHT}: lines 6, 9-12 distinguish SIGHT from the previous approach to group testing on hypergraphs~\cite{chen2007competitive}, on which SIGHT is based. Open source Matlab code for SIGHT is posted online \cite{SIGHTcode}.


\subsubsection{Computational Complexity of SIGHT}

The time complexity of $isDefective(S)$ is application-dependent, so computational complexity is here defined as a function of the number of tests required (\textit{i.e.}, calls to $isDefective(S)$). Each call to $BinSearchSIGHT$ takes no more than $\lceil \log_2(a_0) \rceil$ tests to find an element of the defective set, and this operation is performed at most $k_{max}$ times before either a defective set is found or the algorithm aborts. Additionally, all subsets of $D$ larger than $k_{min}$ must be tested to ensure the defective set returned is minimal. The resulting upper-bound on the number of required tests by SIGHT is thus:

\begin{align} \label{t_max_SIGHT}
    max(\#tests|SIGHT) = k_{max}\lceil\log_2(a_0)\rceil+\sum_{j=k_{min}}^{k_{max}} {k_{max} \choose j}+1
\end{align}

In practice, this worst case is rarely realized. Since $a_0 \le N$, each run of SIGHT requires $O(log N)$ tests.

However, not all tests are necessarily equal. \textit{E.g.}, in the power systems application studied here, tests of defective sets require much more computation time than tests of non-defective sets, as shown in Sec. \ref{sec:results}.  The required number of tests of defective sets per each run of SIGHT is upper-bounded by: 

\begin{align}\label{eq:maxposSIGHT}
    max(\#defective Tests | SIGHT) = k_{max} \lceil \log_2 (a_0)\rceil
\end{align}
Note that this is explicitly a function of $k_{max}$.

\subsection{Random Chemistry} \label{sec:RC}

\subsubsection{Inspiration for RC}
The basic idea for the RC algorithm was originally proposed by Kauffman \cite{kauffman1996home} as a hypothetical method for identifying minimal auto-catalytic sets of interacting molecules from within a very large set of molecules. He suggested testing random half-sets until the products of auto-catalysis were detected in one subset, and repeating this halving process until a minimal auto-catalytic set was discovered (hence the moniker ``Random Chemistry''). Although not previously presented as such, this is an adaptive group testing method.

Inspired by the RC idea, an RC algorithm was implemented for finding genetic interactions between single nucleotide polymorphims that predispose for disease \cite{eppstein2007genomic}. Later, a version of RC was implemented for finding a small set of transmission line outages in electric power networks that trigger cascading power failure, requiring only $O(log N)$ tests per successful run~\cite{eppstein2012random}. 

\subsubsection{RC Algorithm}
As in SIGHT, RC takes as input the set of nodes $V$ and returns either a single minimal defective $k$-set (for $k_{min} \leq k \leq k_{max}$), or the empty set (if the algorithm aborts). The bounds on $k$ are specified for RC exactly as they are in SIGHT. Pseudocode for RC, as implemented in this study, is provided in Algorithm \ref{alg:RC}.

\begin{algorithm}
\caption{RC$(V,A,k_{min},k_{max},t_{max})$} \label{alg:RC}
    \begin{algorithmic}[1]
    \State $a_0\gets A(0)$
    \State $S\gets SAMPLE(V,a_0)$
    \If {$\neg \, isDefective(S)$}
        \State \textbf{return} $\emptyset$
    \EndIf
    
    \For {$i \gets 1, |A|$}
        \State $a_{i} \gets A(i)$
        \State $t = 0$
        \State $flag \gets false$
        \While {$t < t_{max} \land \neg flag$}
            \State $t \gets t + 1$
            \State $S_{new} \gets SAMPLE(S,a_{i})$
            \If {$isDefective(S_{new})$}
                \State $S \gets S_{new}$
                \State $flag \gets true$
            \EndIf
        \EndWhile
        \If {$\neg flag$}
            \State \textbf{return} $\emptyset$
        \EndIf
    \EndFor
    \State \textbf{return} $bottomUpRC(S,k_{min},k_{max})$
    \end{algorithmic}
\end{algorithm}
~\\

RC requires more control parameters than SIGHT. Specifically, RC requires not only the initial set size $a_0$, but an entire subset reduction scheme $A=\{a_0,a_1,\dots ,a_{final}\}$, such that $N \geq a_0 > \dots > a_{final} \geq k_{max}$, as well as the maximum number of tries $t_{max}$ to find a defective set at each subset size $a_i$ before aborting (although we use constant $t_{max}$, one could conceivably select a different $t_{max}$ for each subset size).

As in SIGHT, RC begins by drawing a random sample $S$ of elements from $V$ such that $|S|=a_0$ (Alg. \ref{alg:RC}: lines 1-2) and aborts, returning the empty set, if $S$ does not contain a defective set (Alg. \ref{alg:RC}: lines 3-5). If an initial defective set $S$ of size $a_0$ is found, subset reduction proceeds according to the reduction scheme $A$ (loop starting at Alg. \ref{alg:RC}: line 6).

The sampling loop in RC (Alg. \ref{alg:RC}: lines 10-17) stochastically attempts to find a defective subset $S_{new}$ of size $a_{i}$, from set $S$ of size $a_{i-1}$. If no such subset is found after $t_{max}$ attempts, the algorithm aborts and returns the empty set (Alg. \ref{alg:RC}: lines 18-20). When a subset of size $a_{final}$ is found that causes a cascade, a bottom-up search is conducted (Alg. \ref{alg:RC}: line 22, which calls subroutine $bottomUpRC(S,k_{min},k_{max})$, pseudocode not shown), testing all subsets of size $k$, for $k=k_{min},\dots ,k_{max}$ (in random order for each $k$), returning either the first defective $k$-set found or the empty set, if no minimal defective set of size $\leq k_{max}$ exists in $S$. The subroutine $bottomUpRC$ differs slightly from $bottomUpSIGHT$, in that $bottomUpRC$ must test all subsets of $S$ until a defective subset is found or $k>k_{max}$. In SIGHT, some subsets of $S$ have already been tested during the binary search, so $bottomUpSIGHT$ only needs to test the subsets of $S$ that have not been previously tested. Open source Matlab code for RC, as implemented in this study, is posted online \cite{RCcode}.

\subsubsection{Computational Complexity of RC}
The RC algorithm requires a test of the initial subset, plus up to $t_{max}$ tests for each of the $|A|-1$ reduction steps, plus additional tests for the bottom-up search of the set of size $a_{final}$. Thus, the maximum potential number of tests required by an RC run is:

\begin{align} \label{RC_big_O}
max(\#tests|RC) = 1+(|A|-1)t_{max} + \sum_{k=k_{min}}^{k_{max}} {a_{final} \choose k}
\end{align}

In practice, (using the RC parameters specified in the experiments presented here) only a few tests ($\ll t_{max}$) are typically required during each stage of the subset reduction, and $2$-sets are found most frequently (Section \ref{sec:results}) even when $k_{max}>2$, so the average number of tests required is much lower than this (Section \ref{sec:results}). As long as $a_{i}=a_{i-1}/c$, for some constant(s) $c > 1$, then $|A| \propto log(a_0)$, for some $a_0 \le N$. Under these circumstances, each RC run requires $O(log N)$ tests.


In RC, each reduction step requires exactly one defective test, so the number of defective tests required by a successful RC run is constant. For a reduction scheme of length $|A|$ followed by a brute force search until one defective set is found, the maximum number of tests of defective sets required by an RC run is: 
\begin{align}\label{eq:maxposRC}
    max(\#defective Tests | RC) = {|A|+1} 
\end{align}

Assuming set sizes in $A$ are reduced by a minimum constant fraction $c$, \ref{eq:maxposRC} reduces to:
\begin{align}
    max(\#defective Tests | RC) \le \lceil \log_c (a_0)\rceil + 1
\end{align}

Thus, the maximum number of defective tests is not a function of $k_{max}$, as it is in SIGHT (compare to Eq. (\ref{eq:maxposSIGHT})). Given that there is some unknown probability of aborting during any reduction step, the average number is less than this. 

\subsection{Determining Control Parameters} \label{sec:alg_opt}

Good performance of both SIGHT and RC is contingent on the selection of good control parameters.  In the power systems application, these may vary with the particular data set, power systems simulator, the criteria for determining cascading failure, and $k_{max}$.

Since SIGHT has just the single control parameter $a_0$, it is relatively straightforward to find a good value by performing a parameter sweep on a sample of runs (as in Section \ref{sec:paramsweep}). Other than the stochasticity induced by the random initialization of the initial subset of size $a_0$, SIGHT is completely deterministic with no additional control parameters to specify.

RC also requires the user to determine $a_0$, and it is interesting to note that the values of $a_0$ that minimize the number of tests are not necessarily the same for RC and SIGHT, even when applied to the same problem (see Section \ref{sec:results}). In addition, RC requires the user to specify the entire subset reduction scheme $A$ (or, more commonly, rules for implicitly determining $A$, starting from $a_0$), as well as the maximum number of tests RC allows at each given set size $t_{max}$  before aborting (Alg. \ref{alg:RC}: line 10). Kauffman \cite{kauffman1996home} described his hypothetical RC set reduction scheme $A$ with $a_0 = N, a_i = a_{i-1}/2$, and $t_{max} = \infty$, but these are not necessarily the optimal parameter settings for a given application.

If there is exactly one minimal defective $k$-set, and the cost of each test is constant, then one can derive an optimal set reduction scheme $A$ for RC ~\cite{buzas2013optimized}. However, in experimentation with power systems applications, where there are an enormous number of minimal defective $k$-sets and the cost of each test is variable, we have observed that the best performing set reduction scheme $A$ is problem-dependent.

Through empirical experimentation (unpublished), effective heuristics for selecting RC parameters were reported in previous work \cite{eppstein2012random}. Specifically, for a model of the Polish power grid ($N = 2,896$) at peak winter load \cite{zimmerman2011matpower}, using the power systems simulator DCSIMSEP \cite{Hines2016cascadingChapter}, RC was defined using $a_i = \lceil a_{i-1}/c \rceil$ where $c=2$ for $a_{i-1}>20$ and $c=1.5$ for $20 \ge a_{i-1} > a_{final}$, with $a_{final}=5$ and $t_{max}=20$ random tries per set size~\cite{eppstein2012random}. The initial set size of $a_0 = 80$ was selected so that it required only a few tries to find an initial defective set. This resulted in the set reduction scheme of $A = \{80, 40, 20, 14, 10, 7, 5\}$. After a defective 5-set was found, a brute force search was then performed to find a minimal defective subset of size $\leq k_{max}$.  When applying RC to a larger model ($N=12,706$) of the Western United States (described in Section \ref{sec:test_case}), these same heuristics were adopted with the exception that $a_0$ was raised to 320 to increase the frequency of defective initial sets on this larger test case, so that $A=\{320, 160, 80, 40, 20, 14, 10, 7, 5\}$ \cite{clarfeld2019risk}.

Although these RC heuristics have been found to perform satisfactorily on several test cases (subject to a good selection of $a_0$), it is not feasible to formally optimize them, due to the vast number of possible parameter combinations and the sensitivity of these parameters to specifics of each test case. Nonetheless, as these heuristics represent the current state-of-the-art for RC, we adopt them here, and in Section \ref{sec:paramsweep} we restrict our parameter sweep for RC to finding a good initial subset size $a_0$. It is likely that additional tuning of the other parameters in RC could produce even faster results, but that is beyond the scope of this paper.

\subsection{Experiments} \label{sec:experiments}

\subsubsection{Western US Power Systems Test Case} \label{sec:test_case}
We consider a power systems application where the set of nodes $V$ that are subject to failure comprise all transmission lines (a.k.a. ``branches'') in a power system. 

\begin{figure}[!t]
\centering
\includegraphics[width=3.5in, height=2.5in]{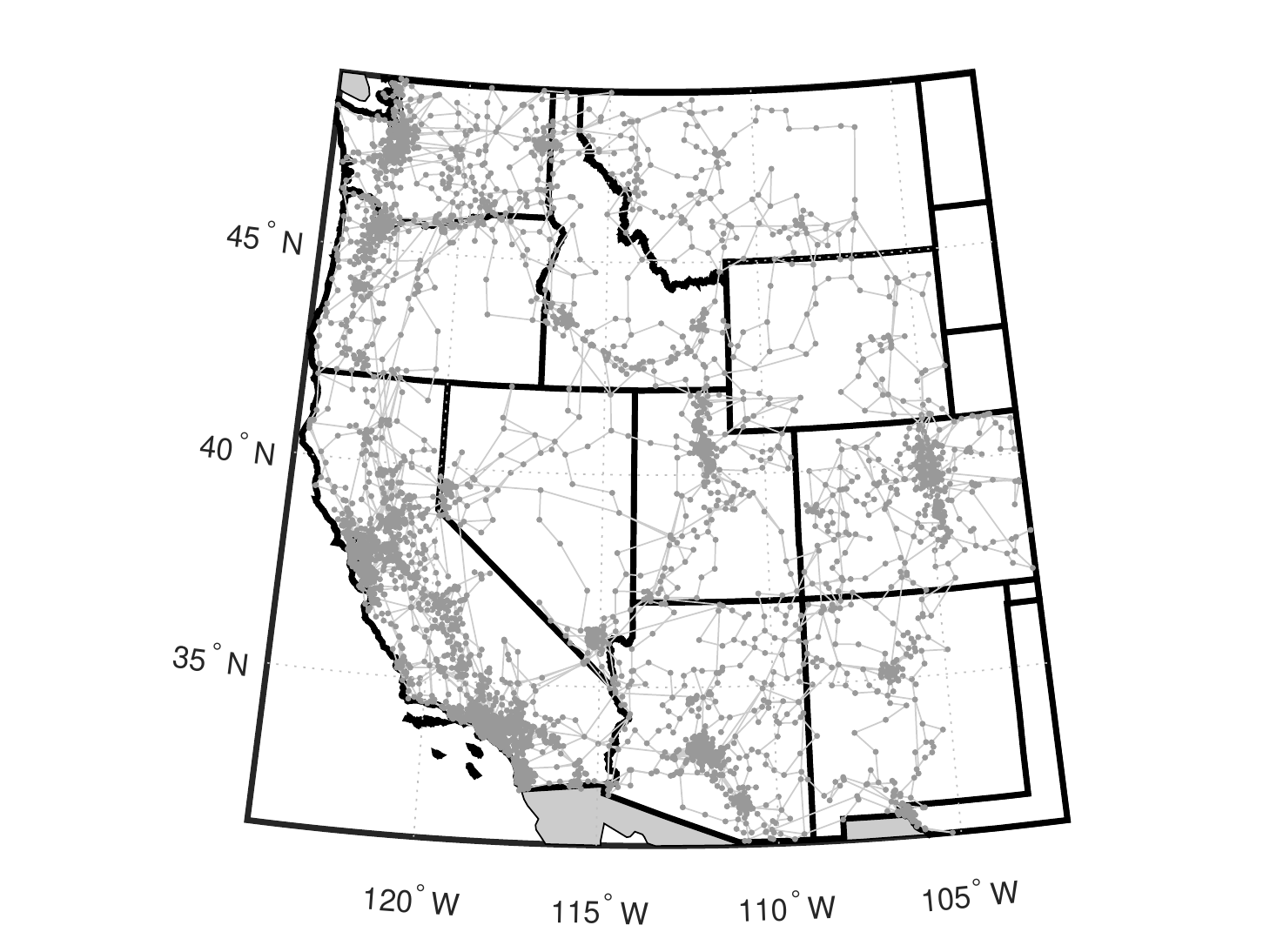}
\caption{Geographic layout of the synthetic 10,000 bus Western US test case.}
\label{western_US_grid}
\end{figure}

As in \cite{rezaei2014estimating, rezaei2015rapid, clarfeld2018assessing, clarfeld2019risk}, a cascading blackout is considered to have occurred when 5\% or more of the total load is shed in DCSIMSEP, a simulator of cascading outages in power systems~\cite{Hines2016cascadingChapter}. Once this threshold is exceeded, a simulation is terminated. The specific test case considered comprises a  large synthetic power system, with a realistic geographical topography based on the footprint of the western Unites States transmission system, which is included in the Electric Grid Test Case Repository~\cite{birchfield2017grid} and has been used in a recent study of risk of cascading failure \cite{clarfeld2018assessing}. The test case contains 10,000 buses (connection points, typically substations, through which generators provide power and loads draw power from the network) and $N=12,706$ branches (transmission lines and transformers) and covers 11 US states (Fig.~\ref{western_US_grid}). The test case was modified to ensure that no branches were initially overloaded and that no single component failure would trigger a cascade, using the adjustments described in~\cite{rezaei2015rapid}. Since short-term (``RateB'') and long-term  (``RateC'') emergency flow limits were not available, they were synthesized using the procedures in \cite{clarfeld2018assessing}.

It is important not to confuse the graph that represents the power grid (as shown in Fig. \ref{western_US_grid}) with the hypergraph being searched; in this test case, ``nodes'' in the hypergraph being searched represent ``edges'' (transmission lines) in the graph that represents the power grid, and ``hyperedges'' in the graph being searched represent sets of transmission line outages.\par

\subsubsection{Determining $a_0$ for the Test Case}\label{sec:paramsweep}

A parameter sweep was performed for a variety of initial set sizes $a_0 \in \{48,64,80,96,112,128,144,160,176,192\}$, using the Western US power systems test case. The same random sets of initial failures (transmission line outages) were chosen for paired trials of SIGHT and RC, for a total of 290,000 trials each: 40,000 trials for each $a_0 \leq 64$; 30,000 trials for each $80 \leq a_0 \leq 144$; and 20,000 trials for each $a_0 \geq 160$. A greater number of trials was required for lower $a_0$, due to the diminishing rate of successful trials as $a_0$ decreases.  All experiments used $k_{min}=2$ and $k_{max} \in \{2,3,4\}$.

After specifying $a_0$, each subset reduction scheme $A$ for RC was created using the same heuristics that have previously been shown effective in \cite{eppstein2012random}, as described in Section \ref{sec:parameterization}. The $a_0$ values for SIGHT and corresponding subset reduction schemes for RC are summarized in Table \ref{tab:params}. 

\begin{table}[!h]
\renewcommand{\arraystretch}{1.3} 
\setlength{\tabcolsep}{10pt} 
\caption{Values tested for $a_0$ for SIGHT and $A$ for RC, in the Western US power systems test case. }
\label{tab:params}
\begin{center}
\rowcolors{1}{white}{light-gray}
\begin{tabular}{ ccc } 
\hline
   SIGHT ($a_0$) & RC ($A$) \\ \hline
  48 & $\{48,24,16,11,8,5\}$ \\
  64 & $\{64,32,21,14,10,7,5\}$ \\
  80 & $\{80,40,20,14,10,7,5\}$ \\
  96 & $\{96,48,24,16,11,8,5\}$ \\
  112 & $\{112,56,28,14,10,7,5\}$ \\
  128 & $\{128,64,32,21,14,10,7,5\}$ \\
  144 & $\{144,72,36,24,16,11,8,5\}$ \\
  160 & $\{160,80,40,20,14,10,7,5\}$ \\
  176 & $\{176,88,44,22,15,10,7,5\}$ \\
  192 & $\{192,96,48,24,16,11,8,5\}$ \\
  \hline
\end{tabular}
\end{center}
\end{table}

The maximum number of tests ($t_{max}$) allowed per RC set size was set to 20, as in \cite{eppstein2012random}. The number of total tests, defective tests, non-defective tests, and elapsed run times needed to successfully find the next minimal defective $k$-set were recorded separately for each of $k_{max} \in \{2,3,4\}$. By recording these metrics until the next successful ``find'', the cost of aborted runs was effectively amortized into these test counts and run times. All experiments were performed in batches of 100 trials, distributed across a high-performance computing cluster. Although clock times on randomly assigned processors in the cluster may vary slightly \cite{VACC}, by performing the experiments in small batches it is assumed that both algorithms were run on sets of processors with similar average clock times. 

\subsubsection{Comparing RC and SIGHT for $a_0=96$}
After determining that $a_0=96$ minimized run times in both RC and SIGHT (see Section \ref{sec:parameterization}), we then ran an additional set of 500,000 paired trials (\textit{e.g.}, starting from the same random initial sets) for each algorithm using $a_0=96$. All other details of these experiments are as described above in Section \ref{sec:paramsweep}.

It is known that, in this power systems application, the distribution of frequencies with which individual branches occur in minimal $k$-sets is long-tailed (approximately power law distributed)\cite{eppstein2012random, clarfeld2018assessing}. It has also been shown that RC systematically under-samples minimal $k$-sets that include the branches that occur most frequently in minimal defective $k$-sets \cite{clarfeld2018assessing}. To assess whether SIGHT contains similar bias, we compare the relative frequencies with which branches occur in the resulting collection of minimal defective $2$-sets found by both RC and SIGHT to the known frequencies of branches found in all minimal defective $2$-sets for this test case (as previously identified in \cite{clarfeld2018assessing}). 

\section{Results} \label{sec:results}

\subsection{Determining $a_0$ for the Test Case} \label{sec:parameterization}

\begin{figure}[!t]
\centering
\includegraphics[width=3.5in]{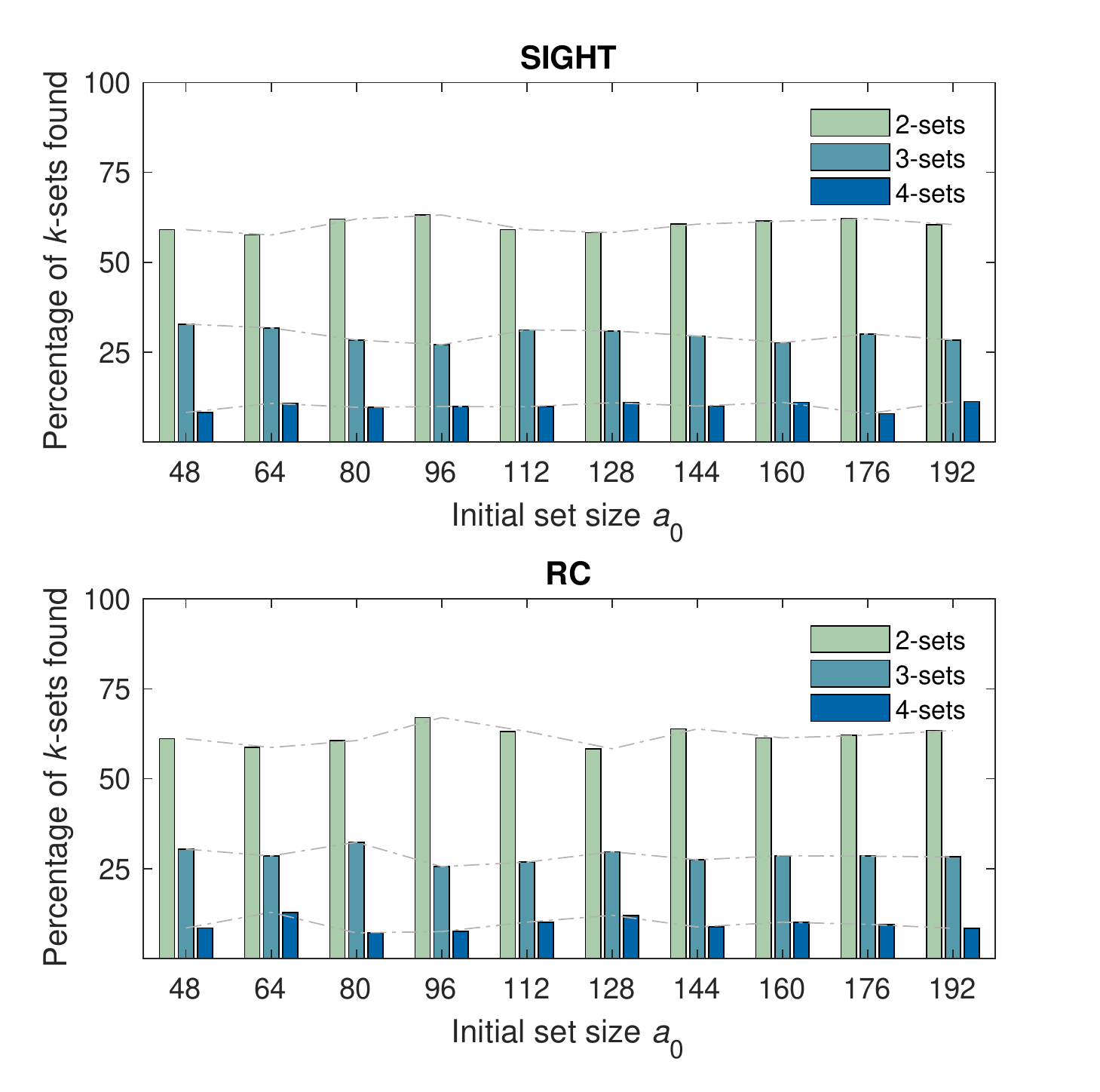}
\caption{The number of $k$-sets of size $k \in \{2,3,4\}$ found in 290,000 paired runs of SIGHT (top) and RC (bottom) with $k_{max}=4$, for $a_0 \in \{48,64,80,96,112,128,144,160,176,192\}$. }
\label{fig:defSetSize}
\end{figure}

RC and SIGHT were similarly effective in finding minimal defective $k$-sets. For example, when $k_{max}=4$, RC and SIGHT found similar proportions of minimal defective 2-sets, 3-sets, and 4-sets, in approximately a 6:3:1 ratio, and this ratio does not appear to be a function of the initial set size $a_0$ (Fig. \ref{fig:defSetSize}). 

\begin{figure}[!t]
\centering
\includegraphics[width=3.5in]{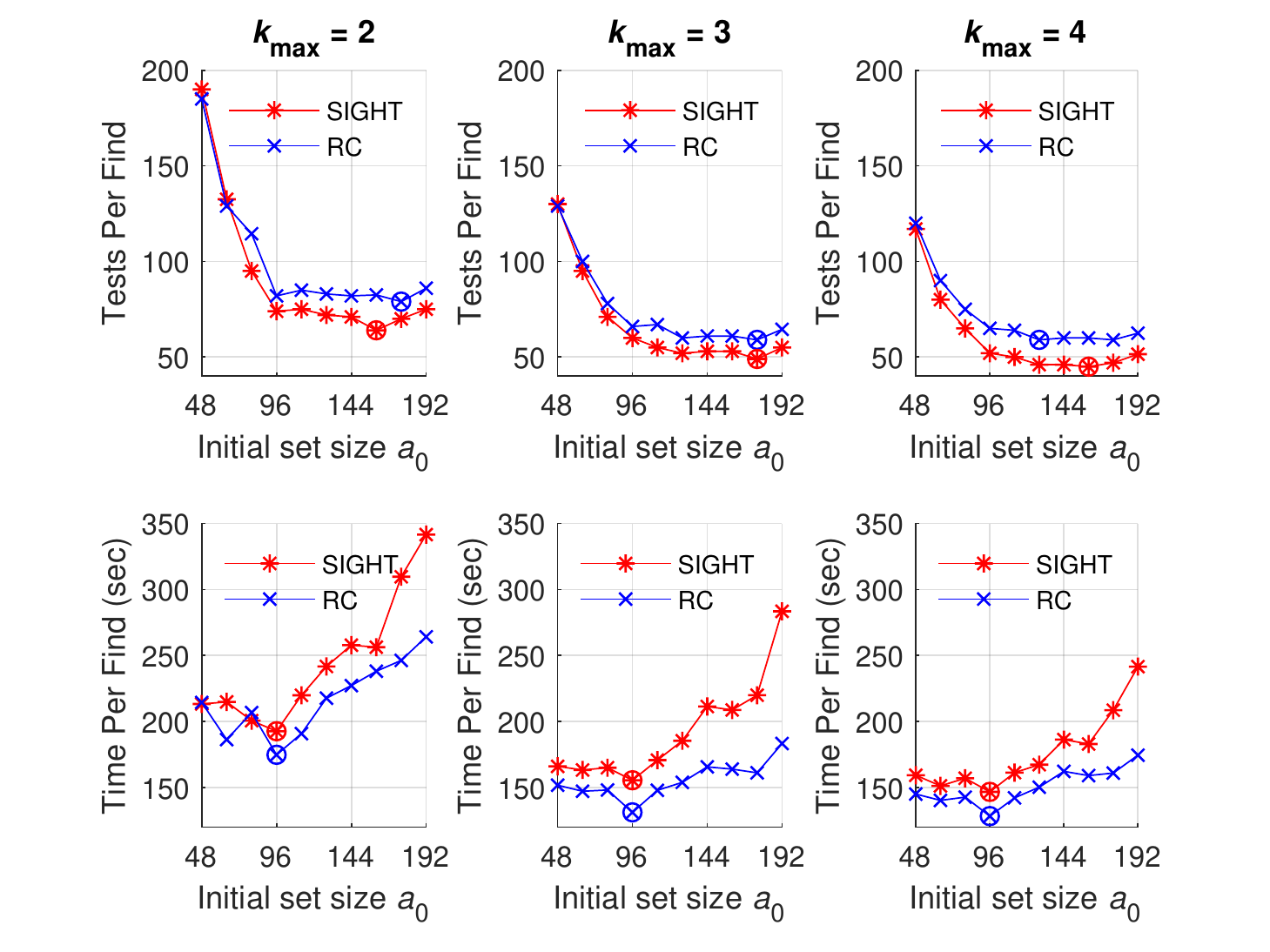}
\caption{The median number of tests (top row) and median run time, in seconds (bottom row), per defective set found by RC and SIGHT, for $a_0 \in \{48,64,80,96,112,128,144,160,176,192\}$ stratified by $k_{max} \in \{2,3,4\}$ (columns). The cost due to aborted runs in between successful finds is included in these metrics. The empirically minimum data points for each algorithm are circled.}
\label{fig:tests_median}
\end{figure}

SIGHT required significantly fewer total tests than RC for all initial set sizes tested (Fig. \ref{fig:tests_median}, top row). For both algorithms, the required number of tests decreased rapidly with increasing $a_0$, for $a_0 \le 96$. Although the specific values of $a_0$ that minimized the number of tests varied between the two algorithms and between different values of $k_{max}$ (see circled data points in the top row of Fig. \ref{fig:tests_median}), the number of tests required by both RC and SIGHT was relatively insensitive to the initial set size for $96 \leq a_0 \leq 192$. At their respective minimum values, SIGHT required 30.5\% fewer tests per find with $k_{max}=2$, 20.3\% fewer tests with $k_{max}=3$, and 23.7\% fewer tests with $k_{max}=4$. All three differences were found to be significant (2-sample Kolmogorov-Smirnov test, $p \ll 0.001$).

However, comparative run times show a very different pattern (Fig. \ref{fig:tests_median}, bottom row). Here, run times are seen to be very non-monotonic and quite sensitive to $a_0$. In this case, run times for both RC and SIGHT happened to be minimized at the same initial set size of $a_0 = 96$, for all $k_{max}$. (Although not shown here, we have found other test cases where the value of $a_0$ that minimized run times varied by both the algorithm used and by $k_{max}$.) In all cases RC had slightly lower median run times per find. Specifically, at $a_0=96$, the median run time per successful find was 3.8, 9.9, and 15.9 seconds faster for RC than for SIGHT, for $k_{max} = 2, 3, 4$, respectively. The time differences between RC and SIGHT were not statistically significantly different for $k_{max}=2$, but were significantly different for $k_{max}\in \{3,4\}$ (2-sample Kolmogorov-Smirnov test $p=0.12$, $p \ll 0.001$, and $p=0.006$, respectfully).

Based on these results, we conclude that $a_0=96$ is a reasonable initial set size for both RC and SIGHT on this test problem, as this minimizes run-time and nearly minimizes the number of required tests.

\subsection{Testing Defective \textit{vs.} Non-Defective Sets.} \label{sec:pos_vs_neg}
Why does SIGHT require more run time than RC on this test case, even though it requires significantly fewer test calls? And why do run times increase for both algorithms for $a_0 > 96$, even though the number of tests required is relatively flat for $96 \le a_0 \le 192$? The answers to these questions lie in the facts that, in the power systems application domain: (i) testing a defective set requires much more computation time than testing a non-defective set; (ii) the number of defective sets tested by SIGHT is higher than for RC; and (iii) the number of tests of defective sets increases for both RC and SIGHT as $a_0$ increases. These facts are illustrated below.

For example, Fig. \ref{fig:pos_vs_neg} illustrates the difference in run times for 3000 random test sets: 500 non-defective $k$-sets and 500 (not necessarily minimal) defective $k$-sets, for each of $k \in \{2, 96, 192\}$. Test simulations were performed using DCSIMSEP \cite{Hines2016cascadingChapter} on the Western U.S. test case (Section \ref{sec:test_case}) on a high-performance computing cluster (described in Sec. \ref{sec:experiments}).  Testing defective sets was approximately 29.2, 27.5, and 4.0 times more costly than testing non-defective sets, for $k=2$, $k=96$, and $k=192$, respectively. The size of the $k$-set being tested also has a significant impact on run time. The median run time for testing non-defective sets of size $k=192$ was $\approx 7.6$ times longer than for $k=2$ (2-sample Kolmogorov-Smirnov test, $p \ll 0.001$), whereas testing defective sets was only $\approx 0.05$ times longer for sets of size $k=192$ than sets of size $k=2$ (2-sample Kolmogorov-Smirnov test, $p \ll 0.001$).

\begin{figure}[!t]
\centering
\includegraphics[width=3.5in]{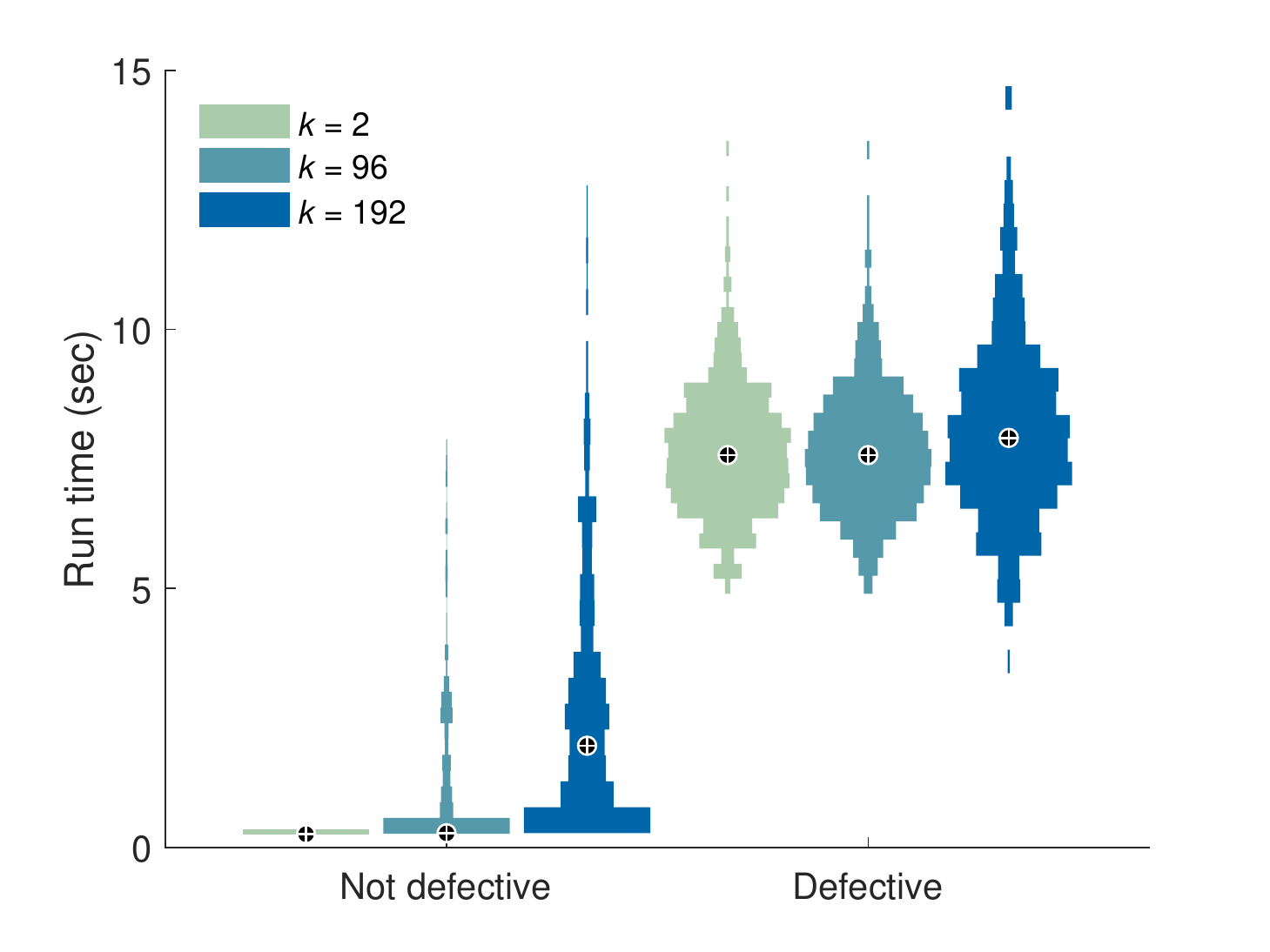}
\caption{Run time required to test 500 random non-defective sets and 500 random defective sets, for each of $k \in \{2, 96, 192\}$, where the test comprises a power systems simulator to test whether a cascading failure occurs. For clarity, medians are marked with crosshairs and each distribution has been independently normalized to the same maximum width. }
\label{fig:pos_vs_neg}
\end{figure}

\begin{figure}[!h]
\centering
\includegraphics[width=3.5 in]{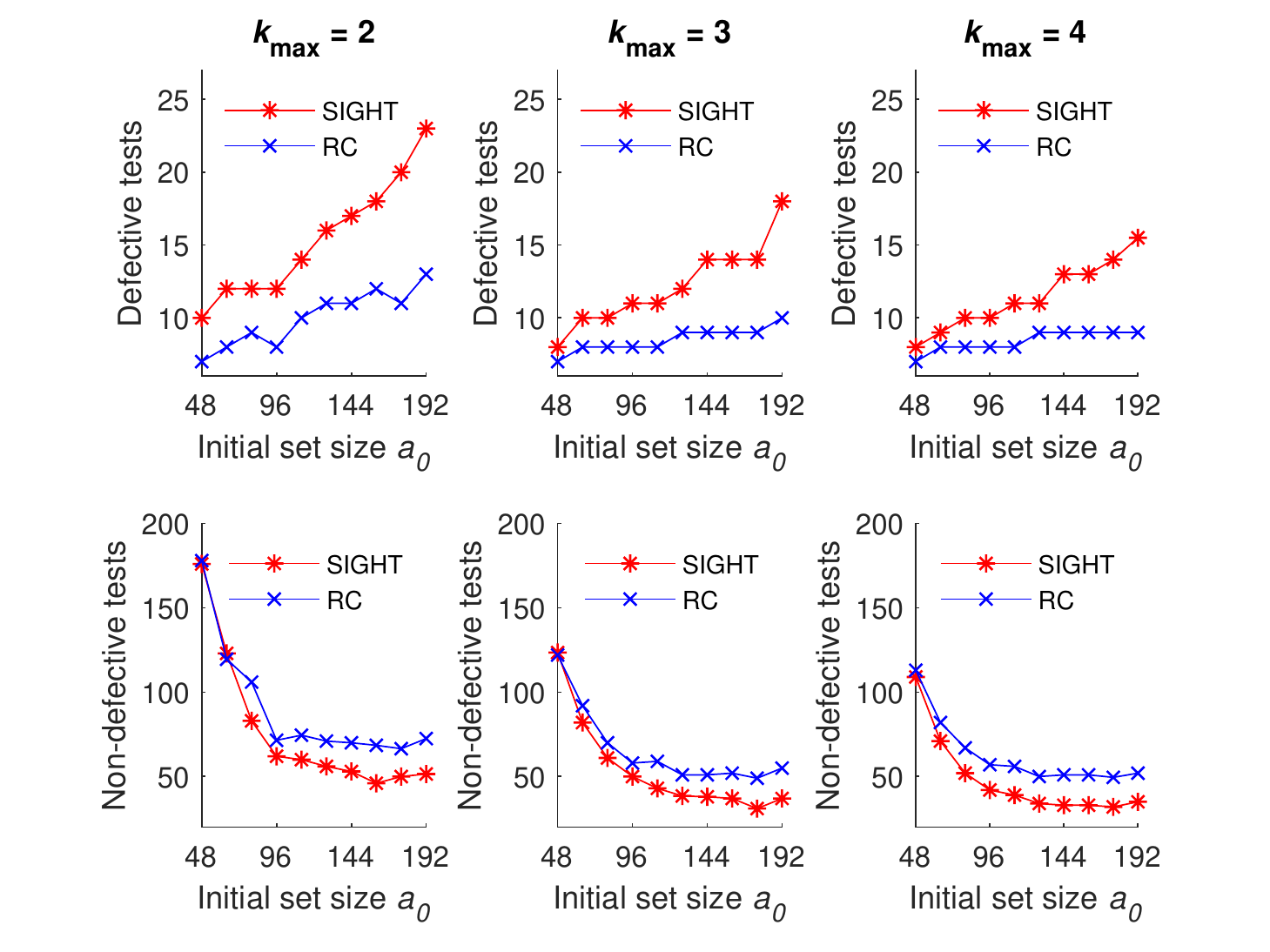}
\caption{The median number of tests of defective sets (top row) and non-defective sets (bottom row), required by RC and SIGHT, per minimal $k$-set found, for initial set size $a_0 \in \{48,64,80,96,112,128,144,160,176,192\}$  stratified by $k_{max} \in \{2,3,4\}$ (columns).}
\label{fig:tests_per_find_p_n}
\end{figure}

 For both RC and SIGHT, the number of tests of defective sets per find increased with increasing $a_0$, while the number of tests of non-defective sets decreased with increasing $a_0$, for all $k_{max}\in \{2,3,4\}$ (Fig. \ref{fig:tests_per_find_p_n}). 
 
 Although SIGHT requires fewer total tests than RC (Fig. \ref{fig:tests_median}, top row), SIGHT requires more tests of defective sets than does RC, and this difference increases with $a_0$ (top row of Fig. \ref{fig:tests_per_find_p_n}).

\subsection{RC \textit{vs.} SIGHT at $a_0=96$} \label{sec:rc_vs_sight}
The results of comparing 500,000 paired trials of RC and SIGHT, with $a_0=96$ and $k_{max}=4$, are summarized in Table \ref{tab:results_overview}. The vast majority (97.2\%) of initial sets tested as non-defective, causing the algorithms to immediately abort; since each algorithm started from the same random initial sets this number is identical for both. The number of subsequently aborted trials was not significantly different for the two methods (Fisher's exact test, $p=0.23$), although RC had slightly fewer subsequently aborted trials than SIGHT (0.62\% \textit{vs.} 0.64\%), resulting a slightly overall higher success rate (2.18\%  \textit{vs.} 2.16\%). 

\begin{table}[!t]
\renewcommand{\arraystretch}{1.3} 
\caption{Overview of results on the Western US Test Case with $a_0=96$ and $k_{max}=4$.}
\label{tab:results_overview}
\begin{center}
\rowcolors{1}{white}{light-gray}
\begin{tabular}{ ccc } 
\hline
   & SIGHT & RC\\ \hline
   Total Trials & 500,000 & 500,000 \\
   Initial Set Aborts & 486,023 & 486,023 \\ 
   Subsequent Aborts & 3,179 & 3,094 \\
   Successful Trial & 10,798 & 10,883 \\ 
   Minimal defective 2-sets found & 6,639 & 6,863 \\
    Minimal defective 3-sets found & 3,099 & 2,896 \\
    Minimal defective 4-sets found & 1,060 & 983 \\
   Median tests per find & 56 & 68 \\
   Median time per find (secs) & 148.5 & 133.8 \\
   Median defective tests per find & 10 & 8 \\
   Median non-defective tests per find & 45 & 60 \\
   
  \hline
\end{tabular}
\end{center}
\end{table}

The distributions of $k$ in the identified minimal defective $k$-sets found were similar to that observed in Fig. \ref{fig:defSetSize}, with a ratio of roughly 6:3:1 for 2-sets, 3-sets, and 4-sets, respectively. Although starting from identical random initial sets, RC and SIGHT only found the same minimal defective sets on 5,085 occasions (approximately 47\% of all paired successful runs for each algorithm), indicating that the initial sets frequently contained multiple minimal defective subsets. Overall, RC required 9.9\% less running time than SIGHT, while SIGHT required 17.6\% fewer tests than RC, per minimal defective set found. RC required 20\% fewer defective tests but 25\% more non-defective tests per find, compared to SIGHT. 

\begin{figure}[!t]
\centering
\includegraphics[width=3.5 in]{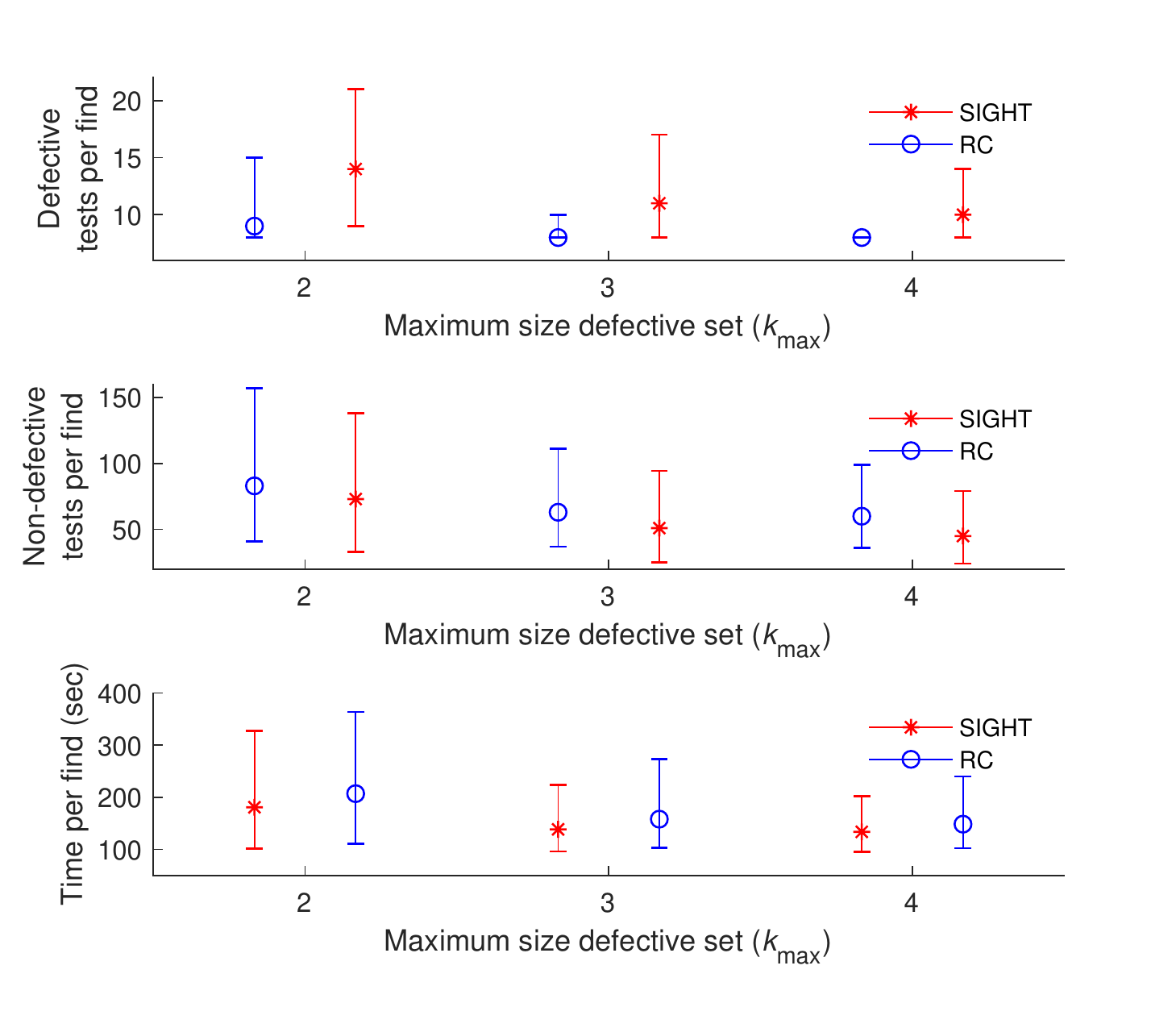}
\caption{The median and interquartile ranges for the number of defective/non-defective tests and running time (in seconds) required by RC and SIGHT, per minimal defective $k$-set found, for $a_0 \in \{96\}$,  stratified by $k_{max} \in \{2,3,4\}$.}
\label{fig:test_time_iqr}
\end{figure}

Run times and number of defective and non-defective tests per find tend to be quite variable for both algorithms (Fig. \ref{fig:test_time_iqr}). Variability decreases as $k_{max}$ increases, because there is a higher chance of aborting and fewer minimal defective sets to be found for smaller $k_{max}$. RC exhibits less variability than SIGHT in the number of defective tests and run time.

\begin{figure}[!h]
\centering
\includegraphics[width=3.5in]{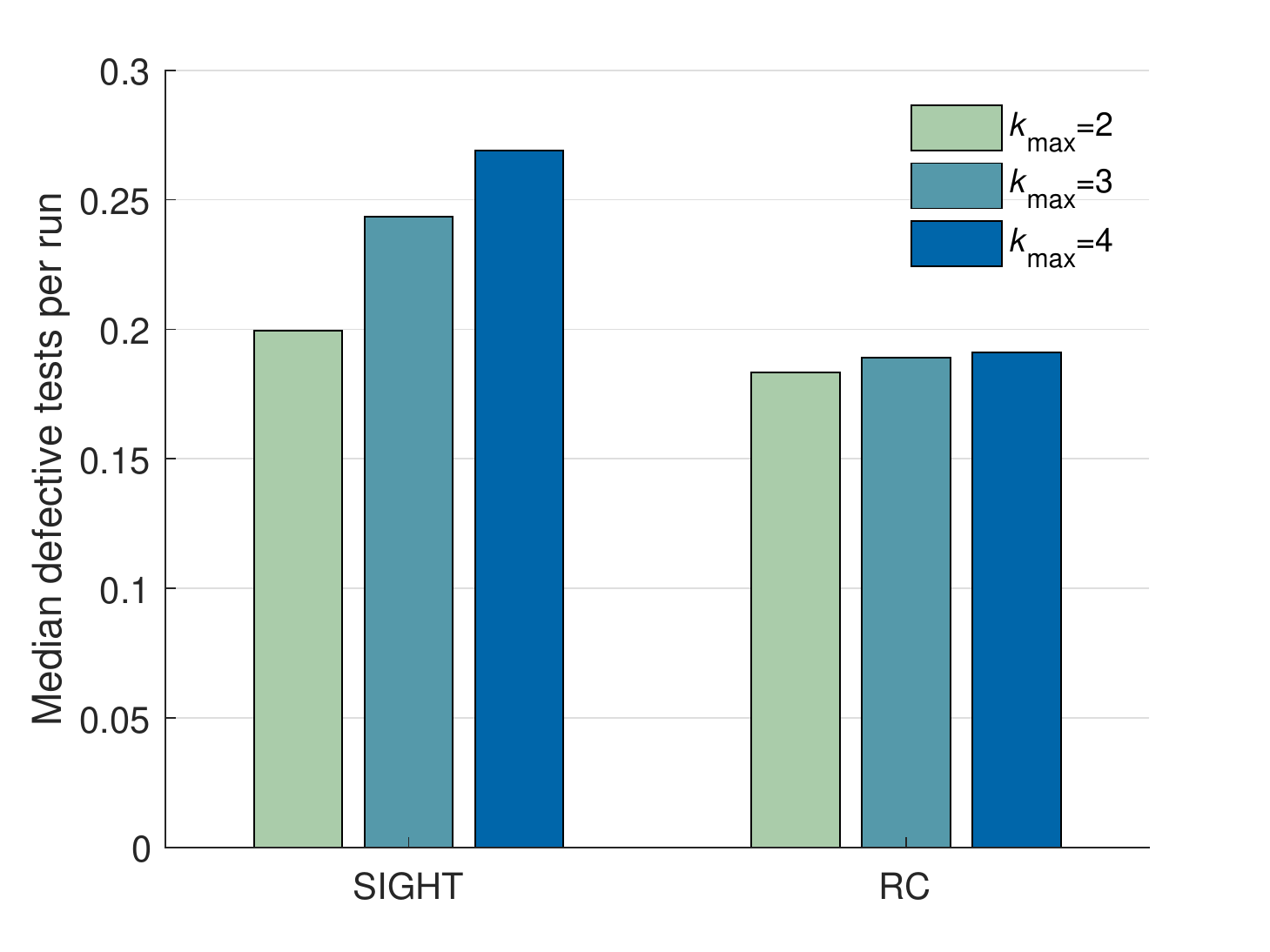}
\caption{The median number of defective tests required by SIGHT and RC per run (including both successful and unsuccesful runs), with $a_0 = 96$, stratified by $k_{max}$. }
\label{fig:posTestsByKmax}
\end{figure}

While Fig. \ref{fig:test_time_iqr} shows that the median number of defective tests required \textit{per successful find} (\textit{i.e.}, factoring in the cost of aborted runs) decreases with increasing $k_{max}$, the opposite is true when considering the number of defective tests required \textit{per run} of each algorithm (Fig. \ref{fig:posTestsByKmax}). Specifically, the number of defective tests required by SIGHT per run is 25.9\% higher when $k_{max}=4$ as compared to to $k_{max}=2$. This is not surprising, since the number of defective tests required by each run of SIGHT increases as a function of $k_{max}$ (Eq. (\ref{eq:maxposSIGHT})).  Although the \textit{maximum} number of defective tests is independent of $k_{max}$ in an RC run (Eq. (\ref{eq:maxposRC})), there is a slight increase in the \textit{median} number of defective tests required by RC runs with increasing $k_{max}$ (Fig. \ref{fig:posTestsByKmax}). This occurs because, as $k_{max}$ increases, fewer RC runs will abort during the $bottomUpRC$ subroutine (due to fewer defective sets having size greater than $k_{max}$ but less than $a_{final}$), and each additional successful RC run caused by an increase in $k_{max}$ will require one additional defective test to be found. 

\begin{figure}[!h]
\centering
\includegraphics[width=3.5in]{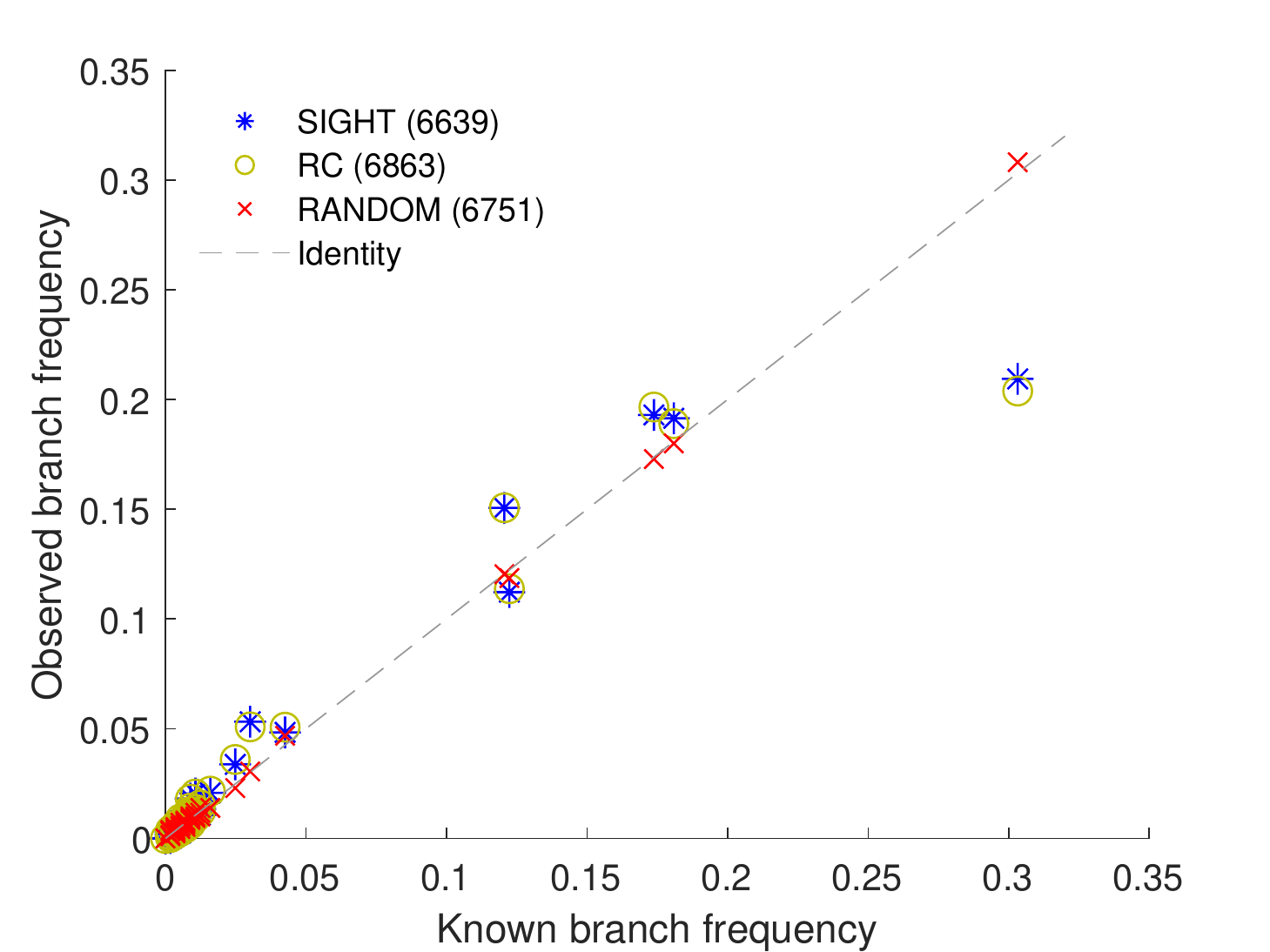}
\caption{Proportion of defective 2-sets identified by SIGHT and RC in which each individual branch appears, as a function of the true proportion of defective 2-sets in which that branch appears for the Western US test case. A comparably sized sample of randomly selected defective 2-sets (RANDOM) was selected to illustrate that very little deviation is expected to occur by chance. Sample sizes are presented in the figure legend. }
\label{fig:biasplot}
\end{figure}

Finally, we observed that the severity and directionality of sampling bias introduced by SIGHT is similar to that introduced by RC, when starting from the same initial sets (Fig. \ref{fig:biasplot}). For example, the most frequently occurring branch is in 30.3\% of all defective 2-sets, but only appears in 20.9\%  and 20.4\% of minimal defective 2-sets (including duplicates) found by SIGHT and RC, respectively (Fig. \ref{fig:biasplot}).

\section{Discussion} \label{sec:discussion}

Since SIGHT only requires a single control parameter $a_0$, it is easier to correctly parameterize than is RC. However, if one adopts the RC parameterization heuristics for set size reduction and for maximum number of tests per set size (albeit not provably optimal) from \cite{eppstein2012random}, then good RC performance can be achieved by also optimizing only $a_0$. 

When RC was first presented for the power systems problem, it was suggested that $a_0$ be selected such that it only requires a few tests to find a defective initial set \cite{eppstein2012random}. Accordingly, when first applying RC to the large Western US Test case, $a_0$ was chosen to be 320 in~\cite{clarfeld2018assessing}.  However, given the results shown here, it is apparent that choosing such a large $a_0$ results in longer run times than necessary.  Counter to our original intuition, run times for this test case are actually minimized at $a_0=96$, even though this causes over 97\% of trials to abort because the initial subset tried is non-defective. This proper parameterization of $a_0$ in RC dramatically reduces the median run time for finding a minimal defective $k$-set in the Western US test case, compared to the value of $a_0=320$ used in \cite{clarfeld2019risk}, with speedups of 2.7, 1.8, and 1.8 for $k_{max} =$ 2, 3, and 4, respectively.

When a trial aborts due to trying an initial subset that is not defective, this has a relatively low computational cost, because simulating non-defective sets in this large power systems problem is much faster than simulating defective sets that cause a large cascading failure. In contrast, trials that abort later in a run require up to $k_{max}\lceil \log_2(a)\rceil+k_{max}$ tests for SIGHT and up to $t_{max}|A|$ tests for RC. The optimal $a_{0}$ finds the right balance between the relative frequencies of these two types of failures.  It is obvious that, as $a_0$ increases, the chances of aborting a trial due to the initial set being non-defective decreases (assuming false negatives are not predominant). It can also be shown that, of runs where the initial sets are defective, the number of aborted runs due to the minimal defective set being too large will increase at a greater rate than the number of successful runs (see Appendix). These competing factors favor a relatively small $a_0$, contrary to the recommendations in \cite{eppstein2012random}. 

Ultimately, it is the combined effects of (i)  the steady increase in the number of defective tests required per find with increasing $a_0$ (top row of Fig. \ref{fig:tests_per_find_p_n}), (ii) the initially steep reduction in non-defective tests required per find as $a_0$ increases, which becomes less steep above $a_0 =96$ (bottom row of Fig. \ref{fig:tests_per_find_p_n}), and (iii) the higher cost of testing defective \textit{vs.} non-defective sets (Fig. \ref{fig:pos_vs_neg}), that together account for the strong non-monotonicity in run times, with a minimum at $a_0=96$ for this test case (bottom row of Fig. \ref{fig:tests_median}). This complex interplay of factors is not possible to predict for a given power systems problem without simulating it. Thus, we recommend doing a parameter sweep to select $a_0$ when applying either RC or SIGHT to a new problem.

\begin{figure}[!h]
\centering
\includegraphics[width=3.5in]{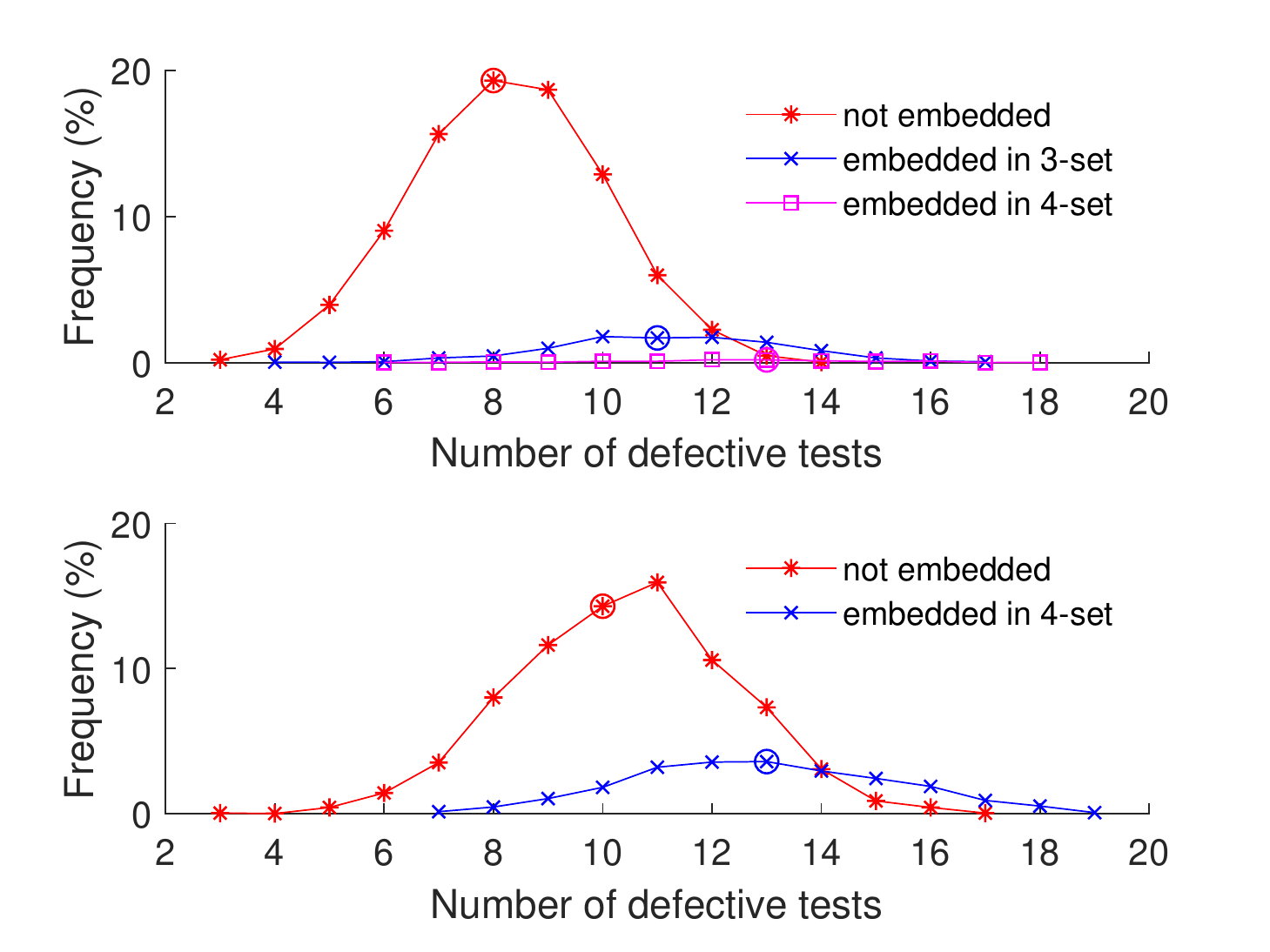}
\caption{Distribution of number of defective tests (for $k_{max}=4$) required per SIGHT trial where a minimal defective $2$-set (top) or $3$-set (bottom) is found in $D$, stratified by whether these were embedded in larger, non-minimal defective sets in $D$. Circled points on the histograms indicate the median number of defective tests for each stratum. }
\label{fig:falseNeg}
\end{figure}

The reliability of tests has a meaningful impact on the computational efficiency of both algorithms, but for different reasons. In SIGHT, a false negative test may cause an element of a minimal defective $k$-set to initially go undetected in the deterministic binary search portion of SIGHT, such that the defective set eventually found in the search is non-minimal. When this occurs, the element subsequently added to the growing set $D$ is not actually part of a minimal defective set.  This is what necessitates the need for the computationally costly final search of $D$ to ensure that the defective set returned is minimal. Figure \ref{fig:falseNeg} shows the frequency with which this happens, and the additional defective tests required. Of all minimal defective 2-sets found by SIGHT, 9.7\% were found embedded in a non-minimal defective 3-set in $D$ and 1.0\% were embedded in a 4-set as a result of false negatives; similarly, of all minimal defective 3-sets found, 22.5\% were found embedded in a non-minimal defective 4-set. 

In contrast, while false negative tests can increase the required number of tests at a given set size for RC, the stochastic nature of subset selection in RC enables it to find a defective set despite the presence of false negatives. The bottom-up search of the final set of size $a_{final}$ is always required by RC, whether or not false negatives are present during the subset reduction steps.

Both algorithms find similar numbers of minimal defective $2$-sets, $3$-sets, and $4$-sets, with a bias toward finding minimal defective $k$-sets with smaller $k$. However, this bias toward smaller $k$ occurs for different reasons in the two algorithms. In RC, with every subset reduction step there is a greater chance of preserving more smaller minimal defective $k$-sets than larger ones, since it is more likely that a larger minimal defective $k$-set will be disrupted during the sub-sampling procedure. Furthermore, the bottom-up search of the set of size $a_{final}$ at the end of RC, will always return one of the smallest contained minimal defective $k$-sets.  In contrast, the binary search step in SIGHT searches for the leftmost element in $S$ that is the rightmost element of a defective $k$-set in $D \concat S$. Thus, the larger a minimal defective set is, the less likely it is to be selected by this procedure. 

The time complexity of successful runs of RC and SIGHT are both $O(log N)$ in the number of tests required. However, this does not mean that their average performances are the same. Overall, in the power systems test case used here, SIGHT required fewer total tests than RC but more tests of defective sets, under nearly all circumstances tested. Furthermore, the maximum number of defective tests per run increases with increasing $k_{max}$ for SIGHT but not for RC. Thus, which algorithm is faster depends on the relative costs of testing defective \textit{vs.} non-defective tests, and in the power systems application this relative cost is problem-specific. For the particular power systems simulator, power systems test case, and cascading failure threshold used in these experiments, RC was slightly faster than SIGHT, for all $a_0$. However, this will not necessarily always be the case.

\begin{figure}[!t]
\centering
\includegraphics[width=3.5in]{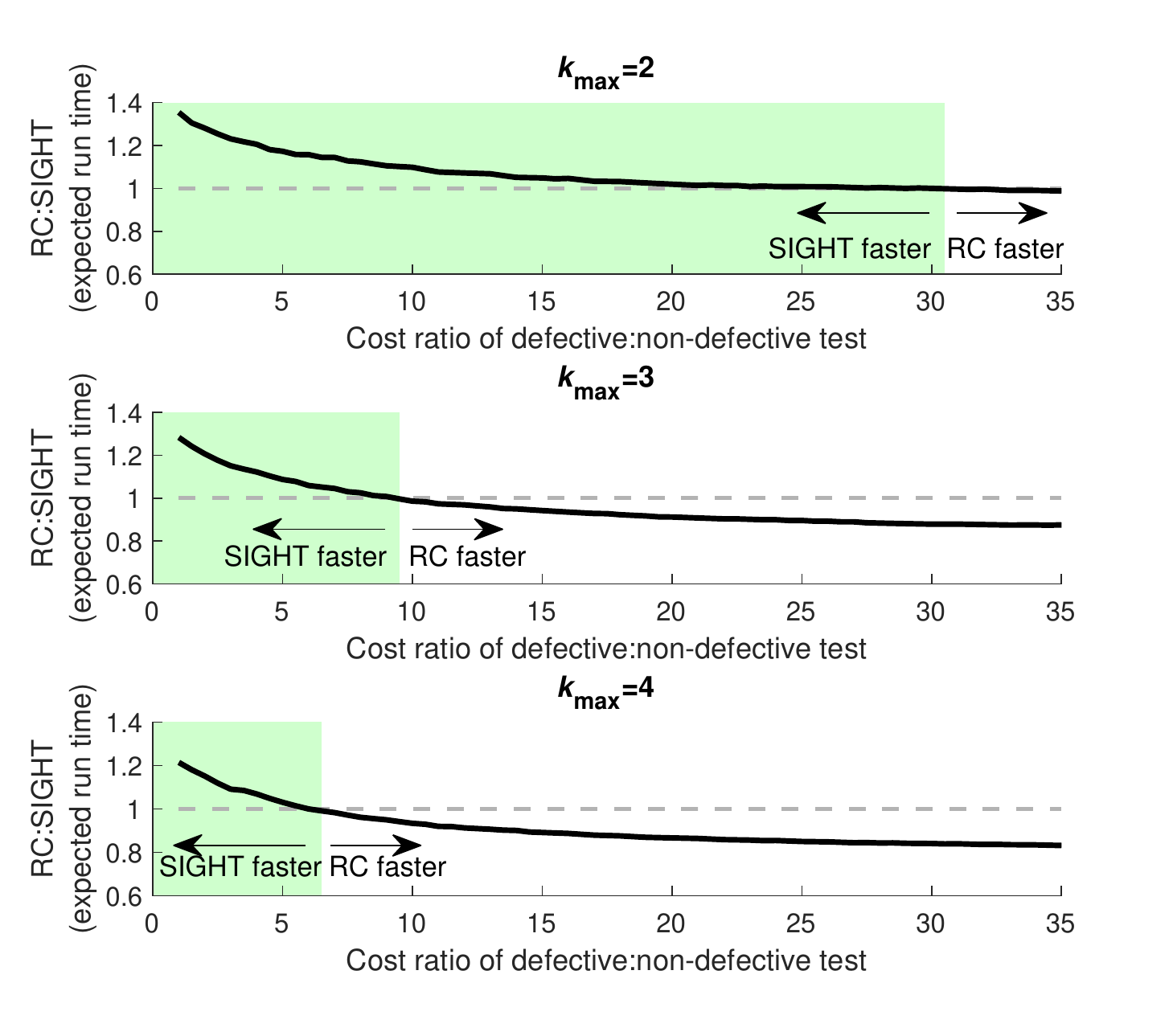}
\caption{The expected ratio of run time for RC:SIGHT as a function of the ratio of the cost of testing defective sets:non-defective sets. The highlighted regions indicate where SIGHT is expected to be faster than RC. Results are shown for $k_{max} \in \{2,3,4\}$.}
\label{fig:costratio}
\end{figure}

For example, using the observed number of defective and non-defective tests required per find at $a_0=96$, we can approximate the expected relative performance of RC and SIGHT as a function of the relative cost of testing  defective \textit{vs.} non-defective sets (if averaged over all set sizes tested) (Fig. \ref{fig:costratio}), where the shaded regions indicate when SIGHT is expected to be faster than RC. Note that SIGHT becomes less efficient relative to RC as $k_{max}$ increases. For $k_{max}=2$, SIGHT is expected to be faster than RC when the cost of defective tests is, on average, up to 33 times greater than the cost of non-defective tests. However, when $k_{max}=4$ we expect that SIGHT will only be faster than RC when the cost of defectives tests is, on average, up to 6.5 times the cost of non-defective tests. However, predicting the average cost ratio for a particular system is not feasible, since the sizes of the sets being tested changes throughout RC/SIGHT, and the relative cost of defective/non-defective tests changes with set size as well as other specifics of a given power system. Thus, simulations are required to determine which algorithm is faster for a given system.

Although we suspect that, due to the high computational cost of simulating cascading failures,  RC will be faster than SIGHT for most realistic problems in this power systems application, not all potential applications of these algorithms have higher costs for testing defective sets. For example, group testing has been proposed as a method for performing feature selection for classification tasks \cite{zhou2014parallel}. In this application, tests would determine whether a set of features achieves a specified level of performance on the classification task. ``Defective'' tests, in which a classifier exceeds the performance threshold, may actually be faster than ``non-defective'' tests, since after the threshold is reached classifier training can be aborted.  Thus, it is expected that SIGHT would be (potentially much) faster than RC in the feature selection application. Group testing has also been employed  as a method for discovering synergistic reactions between drugs \cite{remlinger2006statistical,severyn2011parsimonious,hughes2006pooling,borisy2003systematic}, an application in which the cost of tests is constant; hence SIGHT would be expected to outperform RC for this application (for all $k_{max}$). 

Both RC and SIGHT exhibited a remarkable similarity in the degree and direction of the sampling bias, as shown for 2-sets, when searching the same initial subsets. This occurred even though only 65\% of paired trials found the same minimal defective $2$-sets in SIGHT as in RC, which indicates that most of the bias is likely introduced in the selection of the initial set. For example, the most frequently occurring branch is in 30.3\% of all minimal defective 2-sets, but occurred in only 21.3\% (SIGHT) and 20.6\% (RC) of the random initial defective sets of size 96 in which a minimal defective 2-set was ultimately found. Since the most frequently occurring branch was only slightly less prevalent in the final minimal defective 2-sets found by each algorithm, at 20.9\% (SIGHT) and 20.4\% (RC), this indicates that little additional under-sampling of the most frequent branch occurred when reducing from a defective initial set to an embedded minimal defective 2-set. We believe that this under-representation of the most frequent branch in initial defective sets occurs because of the uniform random selection of the initial sets, in contrast to the highly non-uniform (but \textit{a priori} unknown) prevelance of branches in minimal defective sets, in this power systems example. Any group testing strategy that employs uniform random pooling of individuals for an initial test is thus likely to introduce similar sampling bias, in this application. 

In the power systems application, one of the primary goals of collecting a large sample of minimal defective $k$-sets is to estimate the total number of such sets that exist, as an important step towards estimating the overall system risk of cascading failure \cite{eppstein2012random, rezaei2014estimating, rezaei2015rapid, clarfeld2018assessing}. This estimation is necessary, since it is not possible to find all minimal defective $k$-sets, for $k>2$, in realistically-sized power systems models, due to the combinatorial explosion in the number of these sets. The non-uniform likelihoods of each minimal defective $k$-set being sampled \textit{via} group testing (as discussed in Sec. \ref{sec:rc_vs_sight}) makes estimating the total number of these $k$-sets particularly challenging \cite{clarfeld2018assessing}.  Recently, \cite{clarfeld2018assessing} presented methods  that take advantage of the observed sampling bias in RC to estimate both lower and upper bounds on the total number of minimal defective $3$-sets. These bounding methods will be equally applicable to SIGHT, since it exhibits a sampling bias very similar to that of RC.

\section{Conclusions} \label{sec:conclusion}

Quantifying risk to power transmission networks from cascading failures has remained a challenge for a variety of reasons, including the intractable search space of all possible $k$-sets, the enormous quantity of minimal defective $k$-sets, the heterogeneous distributions of branch occurrences in minimal defective $k$-sets, and the costly simulations needed to identify these sets. This paper describes a new group testing-inspired algorithm (SIGHT) for efficiently sampling defective sets that induce cascades in power transmission networks. The performance of SIGHT is compared to that of RC, a state-of-the-art stochastic group-testing sampling method. 

Since SIGHT requires only one control parameter, it is easier to achieve optimal parameterization than it is for RC. For the test case used herein (using the heuristics recommended in \cite{eppstein2012random} for setting the additional parameters in RC), both SIGHT and RC were found to have the fastest run times using an initial set size of $a_0=96$, even though this resulted in aborting over 97\% of the runs due to the initial set testing as non-defective.  This contradicts previously recommended guidance for choosing an initial set size for RC \cite{eppstein2012random}, where it was suggested to make $a_0$ large enough so that it only requires a few tries to find a defective initial set.

Although RC is still currently the fastest method of finding minimal $k$-sets that cause cascading failures in power systems, our findings indicate that additional optimization of RC parameters may make this algorithm even faster. Increasing $t_{max}$ would increase the number of (cheap) non-defective tests, while decreasing the number of (expensive) aborted runs, and so possibly reduce run times per find. Similarly, increasing the set size reduction factors, while also increasing $t_{max}$, would reduce the ratio of defective to non-defective tests per run, thereby potentially speeding up RC on the power systems application. Although we have yet to find rules for selecting $t_{max}$ or the subset reduction factor(s) that improve upon the heuristics offered in \cite{eppstein2012random}, further work is warranted in this area. 

The mechanisms used by the two algorithms are shown to differ in many ways. However, both RC and SIGHT yielded similar distributions of minimal defective $k$-sets, when searching the same random initial defective sets, and with similar sampling bias, such that the most frequently occurring branch in minimal defective $k$-sets is under-sampled. Since this bias appears to have been largely introduced in the random selection of an initial defective set, we expect that any group testing strategy would exhibit similar bias. In \cite{clarfeld2019risk}, we show how this bias can actually be taken advantage of to estimate upper bounds on the total number of minimal defective $k$-sets.

SIGHT requires fewer total tests than RC, but more tests of defective sets. Thus, which method is faster depends on a number of problem-dependent factors that impact the relative costs of testing defective \textit{vs.} non-defective sets.  While SIGHT proved slightly slower than RC on the large power systems test case used herein, future work will consider other applications where the relative cost of testing defective \textit{vs.} non-defective sets is not so large, to determine on which applications SIGHT may outperform RC. 

Finally, we note that,  prior to this work, the connection had not been made between the rich field of group testing and the development of methods for identifying small sets of power systems elements whose simultaneous failure can initiate large cascading blackouts. The SIGHT algorithm differs from its group testing ancestors \cite{chen2007competitive,chen2011revised} for identifying defective hyperedges in hypergraphs, by allowing the size of minimal defective sets to be arbitrarily large and unknown. This would render adaptive group testing algorithms designed to find all minimal defective sets intractable on large problems, such as the power systems application, due to the sheer number of minimal defective $k$-sets. Hence, we use SIGHT and RC to sample from minimal defective sets, with bounded $k$, rather than identifying all minimal defective $k$-sets. In addition, the possible presence of false negatives (as occur in the power systems application and potentially many other applications) has been largely ignored in the literature on adaptive group testing in graphs/hypergraphs  \cite{li2014pooled,chang2010identification,chang2011pooling}, potentially causing them to fail.  Most importantly, the group testing literature has previously treated all tests as having equal cost. Designing group testing algorithms to simply minimize the number of tests per defective set found, as has been the standard practice in the field \cite{du2000combinatorial}, does not necessarily minimize the run time performance of those algorithms in real-world applications where the cost of tests can vary, such as in the power systems application presented here. Thus, in addition to contributing to the power systems literature, our findings regarding both SIGHT and RC algorithms provide new insights and tools to the group testing community. 

\appendix  
\label{only_apendix}

\section*{Appendix}

\begin{figure}[!h]
\centering
\includegraphics[width=3.5in, height=2.5in]{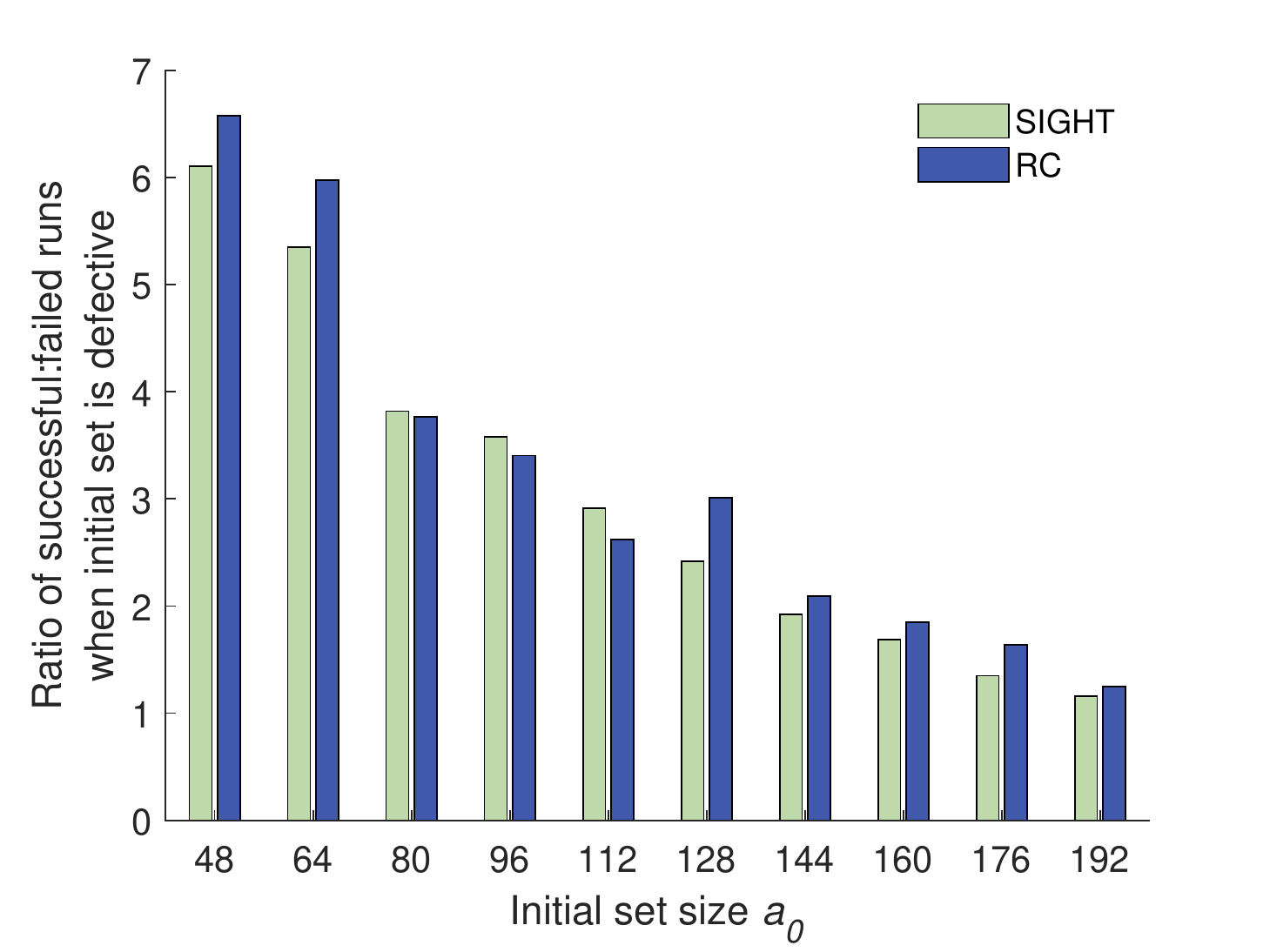}
\caption{The ratio of total runs for SIGHT and RC where a defective $k$-set was found ($k \leq 4$) \textit{vs.} total runs that fail after the initial set of size $a_0$ is defective, for initial set sizes between 48 and 192.}
\label{fig:succ_fail_ratio}
\end{figure}

Empirically, the ratio of successful trials to trials that abort partway through a run (in runs where the initial subset tested as defective) generally decreases with increasing $a_0$, for both algorithms (Fig. \ref{fig:succ_fail_ratio}).

This occurs because the ratio of minimal defective sets of size $(k+1)$:$k$, in some set $S$, increases as the size of the set increases, for all $k$. To prove this, consider a universal set $V$ of elements and the set $\Omega_k$, which contains all minimal defective $k$-sets, for a given $k$. Consider subset $S \subset V$ where the set sizes $|V|=N$ and $|S| = M$. The expected number of minimal defective $k$-sets in $S$ is $|\Omega_k^S| = {M \choose k}/{N \choose k} \times |\Omega_k|$. Now, consider $T \subset V$ where $|T|=M+c$, for some positive constant $c$. Then, it suffices to show: 

$$\frac{|\Omega_{k+1}^S|}{|\Omega_k^S|} < \frac{|\Omega_{k+1}^T|}{|\Omega_k^T|}$$

\begin{proof}
$$
\frac{
    \nicefrac{{M \choose k+1}}{{N \choose k+1}}|\Omega_{k+1}|
        }
    {
    \nicefrac{{M \choose k}}{{N \choose k}}|\Omega_k|
        } < 
\frac{
    \nicefrac{{M+c \choose k+1}}{{N \choose k+1}}|\Omega_{k+1}|
        }
    {
    \nicefrac{{M+c \choose k}}{{N \choose k}}|\Omega_k|
        }
$$\\
$$
{M \choose k+1}{M+c \choose k} < {M \choose k}{M+c \choose k+1} 
$$\\
$$
\frac{M-k}{k+1}{M \choose k}{M+c \choose k} < \frac{M-k+c}{k+1}{M \choose k}{M+c \choose k}
$$\\
$$M-k < M-k+c$$
\end{proof}

Consequently, the number of expected minimal defective sets in $S$ will increase more slowly for $k \leq k_{max}$ (as in a successful run) than for $k > k_{max}$ (as in a run that aborts because the found defective $k$-set is too large). The net effect is that, even though both RC and SIGHT are biased towards finding smaller $k$-sets, the rapid increase in the number of minimal defective sets that are too large, with increasing $a_0$, results in an increase in the proportion of runs that are aborted part way through a run in which the initial set was defective, for both SIGHT and RC. 

\section*{Acknowledgment}

The authors thank Jeffrey S. Buzas and Damin Zhu for helpful discussions regarding group testing algorithms, which ultimately inspired the SIGHT algorithm. We also thank Paul D.H. Hines for generously sharing his knowledge of power systems and inspiring us to work on this problem, providing the code for DCSIMSEP, and for helpful suggestions on the manuscript. The computational resources provided by the University of Vermont Advanced Computing Core (VACC) are gratefully acknowledged. This work was supported in part by NSF Award Nos.~CNS-1735513, ECCS-1254549, and DGE-1144388


\bibliographystyle{unsrt}  
\bibliography{ms}  

\end{document}